\documentclass[useAMS,usegraphicx]{mn2e}
\input epsf.tex

\title[Tilting Orbits]{On the potentially dramatic history of the super-Earth $\rho$~55~Cancri~e} 
\author[Hansen \& Zink] {Bradley M. S. Hansen$^1$\thanks{E-mail:hansen@astro.ucla.edu} \& Jonathon Zink$^{1,2}$\\
$^1$Department of Physics \& Astronomy, University of California Los Angeles, Los Angeles, CA 90095\\
$^2$Department of Physics \& Astronomy, California State University, Northridge, Los Angeles, CA, 91330}

\begin{document}

\date{submitted April 2014}


\maketitle

\label{firstpage}

\begin{abstract}

We demonstrate that tidal evolution of the inner planet (`e') of the system orbiting the star $\rho$~55~Cancri could have led
to passage through two secular resonances with other planets in the system. The consequence of this evolution is excitation
of both the planetary eccentricity and inclination relative to the original orbital plane. The large mass ratio between the
innermost planet and the others means that these excitations can be of substantial amplitude and can have dramatic consequences
for the system organisation. Such evolution can potentially explain the large observed mutual inclination between the innermost
and outermost planets in the system, and implies that tidal heating could have substantially modified the structure of planet~e,
and possibly reduced its mass by Roche lobe overflow. Similar inner secular resonances may be found in many multiple planet
systems and suggest that many of the innermost planets in these systems could have suffered similar evolutions.

\end{abstract}


\section{Introduction}

The importance of secular gravitational interactions in sculpting planetary systems has
long been appreciated as being necessary to understand the regularities of our own
solar system. It is known that 
the long term stability of the terrestrial planets of our solar system is imperilled by
the influence of secular gravitational interactions between Mercury, Venus and
Jupiter (Laskar 1994; Batygin \& Laughlin 2008; Laskar \& Gastineau 2009).
With the discovery of a host of extrasolar planet systems over the past twenty years,
including many multiple systems (Wright et al. 2009; Lissauer et al. 2011a; Fabrycky et al. 2014), the role of
secular interactions in determining the stability and architecture of planetary
systems is becoming of ever greater importance.

The planetary system around the star $\rho$~55~Cancri is of particular importance
in this regard. One of the first systems discovered by radial velocities (Butler et al. 1997),
it also provided one of the first extrasolar giant planets on orbital scales similar to that
of Jupiter (Marcy et al. 2002) and, most recently, the revision of the innermost
planet parameters (Dawson \& Fabrycky 2010) resulted in the photometric detection of a
planetary transit (Winn et al. 2011; Demory et al. 2011) which provides some of the best
constraints on the properties of rocky super-earths (Gillon et al. 2012; Demory et al. 2012).
Furthermore, the presence of a Jovian class planet with an orbital period of 14 days 
(55 Canc~b) places this system in the enigmatic class of `warm Jupiters', which don't fit
naturally into the evolutionary paradigms developed to explain the presence of Jovian
planets with shorter periods. Finally, reports of a tentative detection of an astrometric 
signal with the HST Fine Guidance sensor (McArthur et al. 2004) suggest that this planetary
system may possess a substantial dispersion in orbital inclinations, with a nominal offset
of $37^{\circ} \pm 7^{\circ}$ between the outer planet 55~Canc~d and the edge-on inner planet
55~Canc~e.

It has also been suggested (Kaib, Raymond \& Duncan 2011) that the presence of a distant M-dwarf companion (Mugrauer et al. 2006)
can induce a coherent precession of the entire planetary system (Innanen et al. 1997), potentially leading to 
a substantial obliquity offset between the stellar spin and the orbital plane of the planetary system. However, for
the 55~Cancri~system, spin-orbit coupling between the star and the planets likely quenches such an offset unless
the companion is in a highly eccentric ($e \sim 0.95$) orbit (Boue' \& Fabrycky 2014). Observationally, the situation
is unclear as Bourier \& Henrard (2014) claim to have detected a projected obliquity of $72 \pm 12^{\circ}$, which
Lopez-Morales et al. (2014) claim is impossible given the very low stellar rotation.

All of this recent activity suggests that the 55~Cancri system is interesting for both
its dynamical architecture and history as well as for the composition of the component planets,
in particular 55~Canc~e, as an example of a massive, earth-like planet transitting a bright star.
Our analysis will focus on the fact that 55~Canc~e is
close enough to the star that it should have experienced substantial tidal dissipation,
and consequent orbital evolution. Such evolution can have an effect on more distant
planets too, as secular coupling between the planets can transfer angular momentum
back and forth between the planets in the system (e.g. van Laerhoven \& Greenberg 2012).
In \S~\ref{Tides} we will discuss our model for tidal interactions and describe
the evolution of the system within the context of classical secular theory, and we will
show how this 
can provide a potential explanation for the large inclination dispersion. In \S~\ref{Num}
we investigate the evolution further using numerical integrations, to clarify the dynamical
limitations on the excitation of inclination and discuss the 
influence of the near-resonance of Planets 55~Canc~b and 55~Canc~c on the secular oscillations
of the system. Finally, we discuss the possible influence of tidal heating on the structure
of 55~Canc~e and on the likelihood of other known planetary systems where inclination excitation
could have occurred.

\section{Tidal Evolution in the Classical Secular Approximation}
\label{Tides}

The physical parameters of the innermost planet, 55~Cancri~e, are estimated to be
a mass of $7.99 \pm 0.25 M_{\oplus}$ and a radius of $2.00 \pm 0.14 R_{\oplus}$
(Nelson et al. 2014;  Winn et al. 2011; Demory et al. 2011).
The radius of a perovskite planet of this mass is $\sim 1.9 R_{\oplus}$, using
the fitting formulae of Seager et al. (2007). Thus, although
it is closer in mass to a Neptune or Uranus, this is expected to be a predominantly
rocky planet (although it could have a finite water contribution as well). We wish
to therefore consider a range of tidal dissipation extending from terrestrial values
to ice giant values.

To implement the  tidal dissipation within the secular formalism, we use the same approach as in
Hansen \& Murray (2014), following on from prior work of Wu \& Goldreich (2002) and Greenberg
\& van Laerhoven (2011). We have also previously examined tidal dissipation for extrasolar
 giant gas planets (Hansen 2010), using the formalism of Hut (1981) and Eggleton et al. (1998).
However, the restriction of classical secular theory to the quadratic level in eccentricity
means that we will restrict our tidal expressions to the same order. We must also choose
a dissipation level more appropriate to terrestrial planets. Choosing a bulk dissipation
constant that yields $Q'=10$ for the Earth (e.g. Goldreich \& Soter 1966), we find
\begin{eqnarray}
\frac{d \ln a}{dt}  =  \frac{e^2}{\rm 25\, Myr} \left( \frac{a}{0.1 AU} \right)^{-8} \left( \frac{R_p}{ 2 R_{\oplus}} \right)^{10} \frac{ 8 M_{\oplus}}{M_p} \label{Tideqa} \\
\frac{d \ln e}{dt}  =  \frac{ 1}{\rm 50\, Myr}  \left( \frac{a}{0.1 AU} \right)^{-8}  \left( \frac{R_p}{2 R_{\oplus}} \right)^{10} \frac{ 8 M_{\oplus}}{M_p} \label{Tideq}.
\end{eqnarray}
We have included the estimated mass and radius for 55~Canc~e in these estimates, but let the semi-major axis float. This demonstrates that we expect substantial tidal evolution for a planet like this even out beyond 0.1~AU, unless the strength of the dissipation is several orders of magnitude weaker than expected.

Of course, there is little evolution if the orbit starts close to circular, but the presence of much larger planets in close proximity suggests that a finite eccentricity would be excited by secular perturbations anyway, resulting in tidal evolution (e.g. Mardling \& Lin 2004). Thus, we need to consider the fact that the currently observed configuration could be the result of substantial tidal migration if we wish to infer something about the origin and evolution of this planetary system.

\subsection{Secular Architecture}

Consider then the classical secular solution for the 55~Cancri system with the currently measured parameters of Nelson et al. (2014). In addition to
the classical equations to second order in eccentricity and inclination (Murray \& Dermott 1999), we also include the effects of relativistic precession (Adams \& Laughlin 2006). There is potentially also an
additional precession due to the presence of a distant M dwarf companion,  a 0.26$M_{\odot}$ M dwarf located at 1260~AU (Mugrauer et al. 2006). However, this is too small to affect the dynamics of the planetary configuration on small scales, although it can generate
precession of the system as a unit (Innanen et al. 1997; Kaib et al. 2011; Boue' \& Fabrycky 2014). In order to better match the observations, we
furthermore assume that the planetary system invariable plane is inclined at 37$^{\circ}$ to the line of sight, except for the transitting 55~Canc~e. This increases all the
masses by a factor of 1.25 from that inferred in the radial velocity solution. These are summarised in Table~\ref{ParamTab}.

\begin{table*}
\centering
\begin{minipage}{140mm}
\caption{Basic Architecture of the 55~Cancri Planetary System, based on the parameters
of Nelson et al. (2014) and assuming an inclination of $37^{\circ}$ to the line of sight
for all but the innermost planet. \label{ParamTab}}
\begin{tabular}{@{}lclclc@{}}
Planet & a (AU) & M (M$_J$)\\
\hline
e & 0.035 & 0.0251  \\
b & 0.1134 & 1.056  \\
c & 0.2374 & 0.223  \\
f & 0.774 & 0.184 \\
d & 5.451 & 4.79 \\
\hline
\end{tabular}
\end{minipage}
\end{table*}

The modal
structure for the eigenvalues that results from this configuration is such that the eccentricities of the outer two planets are decoupled from the inner three (as noted by van Laerhoven \& Greenberg 2012), whose interaction will dominate the eccentricity of 55~Canc~e in
the absence of dissipation. On the other hand, all four of the inner planets are coupled to a low frequency inclination eigenmode that represents the precession
of the inner system under the influence of the M dwarf perturber and 55~Canc~d.

\begin{figure}
\includegraphics[width=84mm]{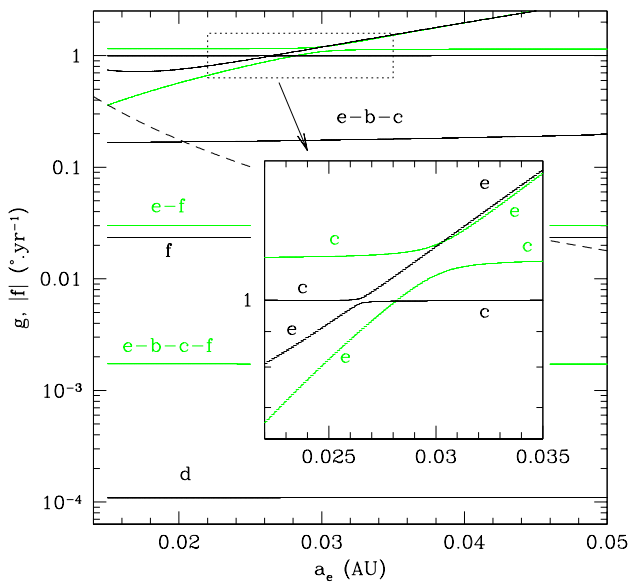}
\caption[afg.ps]{
The black curves show how the precession frequencies of the secular eccentricity eigenvalues $g_i$ change as the
semi-major axis of 55~Canc~e is varied, with all other semi-major axes and masses held constant. The green curves show the equivalent
absolute values of the frequencies for the inclination eigenvalues, $\left| f \right|$. The inset shows a zoom around the most interesting part of parameter space. The various
curves are labelled with the symbols for all planets that make at least a 10\% contribution to the unit eigenvector of that mode.
We see that,
if 55~Canc~e moves through the region $\sim 0.026$--$0.031$~AU, the system will experience secular resonances between both eccentricity
and inclination eigenmodes. In particular, these highlight that the principal interaction is between 55~Canc~e and 55~Canc~c, not the
second most distant 55~Canc~b.  However, all three planets interact via a third eccentricity mode, while the outer two planets are essentially decoupled.
In the case of the inclinations, the inner four planets also precess as a unit under the influence of the outermost planet.
For completeness, the dashed curve shows the value of the relativistic precession frequency at the location of 55~Canc~e.
\label{afg}}
\end{figure}

If the inner planet migrated due to tidal dissipation, then the secular structure of the system will have changed as well.
Figure~\ref{afg} shows how the secular eigenfrequencies of this system vary as we vary the semi-major axis of Planet~e, keeping all masses and
other semi-major axes fixed, from the current location to a semi-major axis of 0.05~AU, corresponding to a period of 4.2~days. The two lowest
eigenfrequencies in eccentricity correspond to the outer planets d and f, and do not change because the outer planets are essentially decoupled.
The lowest frequency inclination eigenfunction corresponds to a mode coupling the four innermost planets.

The higher frequency eigenmodes show a more interesting structure. In particular, we see that the two highest frequency modes in both inclination
and eccentricity experience resonant behaviour in the interval $a_e \sim 0.026$--$0.031$~AU. Thus, if 55~Canc~e starts with a semi-major axis exterior
to 0.032~AU (orbital period of 2.15~days), it will cross first an inclination resonance and, shortly thereafter, an eccentricity resonance. These
will result in mode mixing and can result in interesting behaviour, as we shall see. For the purposes of the following discussion, we
designate the eccentricity resonance as $g_{ce}$ and the inclination resonance as $f_{ce}$.

The dashed line in Figure~\ref{afg} also shows the strength of relativistic precession on these close planetary scales. We have repeated the above
calculation without either of the externally imposed precessions and verified that the behaviour is qualitatively similar. The relativistic
contribution has two quantitative consequences. The first is that the frequencies of the inclination and eccentricity precessions for 55~Canc~e
are the same on small scales without relativity (the black-green pair labelled e in the inset lie on top of one another for $a_e < 0.026$AU)
and the turnup in the eccentricity eigenfrequency of 55~Canc~e at $a_e<0.018$AU is a consequence of relativity.

\subsection{Tidal Evolution}

\begin{figure}
\includegraphics[width=84mm]{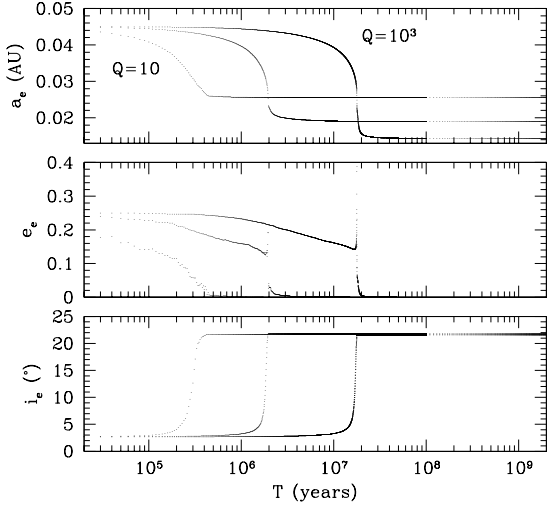}
\caption[ta.ps]{
The upper panel shows the temporal evolution of the semi-major axis of 55~Canc~e as the planetary system evolves under
the influence of tides and secular coupling. We show the evolution for three values of $Q'_e$=10, 100 and 1000. We see that weaker dissipation
actually leads to more orbital migration. The reason for this counter-intuitive behaviour can be found in the middle panel, which shows that the
eccentricity excitation during the resonant crossing is stronger if the tidal dissipation is weaker. The lower panel shows the increase in the
planetary inclination due to the crossing of the inclination resonance.
\label{ta}}
\end{figure}

Let us now consider the tidal evolution of a 55~Cancri analogue in which 55~Canc~e starts with an original semi-major axis $a_e (0) > 0.04$~AU.
The secular coupling of the inner three planets means that the tidal dissipation will remove energy from the system as a whole instead of just
55~Canc~e (Wu \& Goldreich 2002; Greenberg \& van Laerhoven 2011; Hansen \& Murray 2014), and so, the current rather small eccentricities of 55~Canc~b and c may have
been substantially damped. We take the eccentricity of 55~Canc~f to represent a primordial value because it is largely decoupled from the
inner three planets.
 Let us start then, for illustrative purposes, with the same eccentricity for planets b, c and f, $\sim 0.32$. Figure~\ref{ta} shows
the evolution of the system under the influence of tidal dissipation using equations~(\ref{Tideqa}) and (\ref{Tideq}). The modelling of the
interactions between the tides and the secular oscillations is performed using the formalism of Hansen \& Murray (2014). Figure~\ref{ta} shows three
sets of curves, for three different levels of dissipation $Q'_e =10$, 100 and 1000. We see that the consequence of the resonant crossings is the
excitation of inclination and eccentricity for 55~Canc~e. The excitation of the eccentricity also results in a rapid increase in the strength of
the tidal dissipation and a sharp decrease in semi-major axis.

The size of the rapid 'jump' in semi-major axis is dependant on the strength of dissipation $Q'_e$, and is larger for weaker dissipation.
This is a little counter-intuitive but can be understood in terms of the speed at which 55~Canc~e moves through the $g_{ce}$ resonance. The origin
of the high eccentricity of 55~Canc~e is the mixing of the two mode amplitudes during the $g_{ce}$ passage, and a slower passage results in
a larger transfer of amplitude between the two modes. The higher eccentricity means a smaller periastron and more inward migration. 


The increase in inclination $i_e$ seen in Figure~\ref{ta}  offers an explanation for the curious misalignments of 55~Cancri~e and d.
If 55~Canc~e began in the same disk plane as d and all the other planets, but migrated inwards through the secular resonances, then inclination
offsets $\sim 20^{\circ}$ or greater are quite plausible.

\subsection{Analytic Estimates}

Minton \& Malhotra (2011) have investigated the excitation of eccentricity in solar system asteroids by sweeping secular resonances driven by outer planet migration. This formalism can be adapted to understand the excitation of eccentricity and inclination in the 55~Cancri system too. One important difference is that the interaction discussed in MM11 is driven by the change in the frequency of the external resonance, whereas the interaction here is driven by the inward drift of the excited body, as it crosses a fixed external resonance. In appendix~\ref{MM11} we discuss the MM11 analysis,  demonstrate that the end result is similar despite the modification to the model, and present the equivalent expression for the inclination. Here we will briefly review the outline of those results.

MM11 start with a Hamiltonian of the form
\begin{equation}
H_e = -g_0 J + \epsilon \sqrt{2 J} \cos (\bar{\omega}_p - \bar{\omega})
\end{equation}
where $J = \sqrt{a} (1 - \sqrt{1 - e^2})$, $g_0$ is the precession frequency for the longitude of periastron $\bar{\omega}$,
and $\epsilon$ parameterises the strength of the coupling to the externally maintained secular mode, whose orientation is $\bar{\omega}_p$.
To this Hamiltonian, we add the terms that describe the corresponding interaction for the inclination
\begin{equation}
H_i = -h_0 Z + \delta \sqrt{2 Z} \cos (\Omega_p - \Omega)
\end{equation}
and $Z = \sqrt{a (1 - e^2)} (1 - \cos i)$.

For the size of the eccentricity jump from $J_i$ to $J_f$, they derive 
\begin{equation}
J_{f} = J_{i} + \frac{ \pi \epsilon^2}{2 \left|\dot{g_0}\right|} + \epsilon \sqrt{\frac{2 \pi J_{i}}{\left|\dot{g_0}\right|}} \cos \bar{\omega}_{i}.
\end{equation}
The size of the jump is thus regulated by the rate of change of $g_0$ due to tidal dissipation, as seen in the previous section,  as well as the phase of the pericenter precession, $\bar{\omega}_i$. The
expression for this is
\begin{equation}
\dot{g_0} = \frac{n}{8} \frac{\dot{a}}{a} \left( \sum_j \frac{m_j}{m_c} \alpha_j^2 \gamma_j 
 - \frac{15}{2} \frac{G M_c}{c^2 a} \right),
\end{equation}
where $n$ is the orbital frequency, $\alpha_j=a/a_j$ is the ratio of semi-major axes of 55~Canc~e and perturber $j$, 
$\gamma_j =  b^{(1)}_{3/2}+3 \alpha_j \left[ b^{(0)}_{5/2} - 2 \alpha
b^{(1)}_{5/2} + b^{(2)}_{5/2} \right]$ and the
functions $b^{(i)}_s$ are the Lagrange functions (functions of $\alpha_j$).

Taking $\dot{a}/{a}$ from equation~(\ref{Tideqa}) we see that the rate of dissipation actually depends on the eccentricity, which is the
quantity being excited during the transition. Thus, the speed at which the planet crosses the resonance accelerates as the eccentricity
is excited -- a naturally self-limiting process. To estimate the final eccentricity, we thus express
 $\dot{g_0} = e_f^2 g_0/\tau_0$, where $e_f$ is the final eccentricity. In the limit of low eccentricity, $J \sim \frac{1}{2} \sqrt{a} e^2$, so that the final expression is
\begin{equation}
e_f^2 \sim e_i^2 + \frac{\pi \epsilon^2 \tau_0}{2 g_0 \sqrt{a} e_f^2} + \epsilon \sqrt{\frac{2 \pi  \tau_0}{g_0 \sqrt{a}}} \frac{e_i \cos \bar{\omega}}{e_f}. \label{ejump}
\end{equation}

\begin{figure}
\includegraphics[width=84mm]{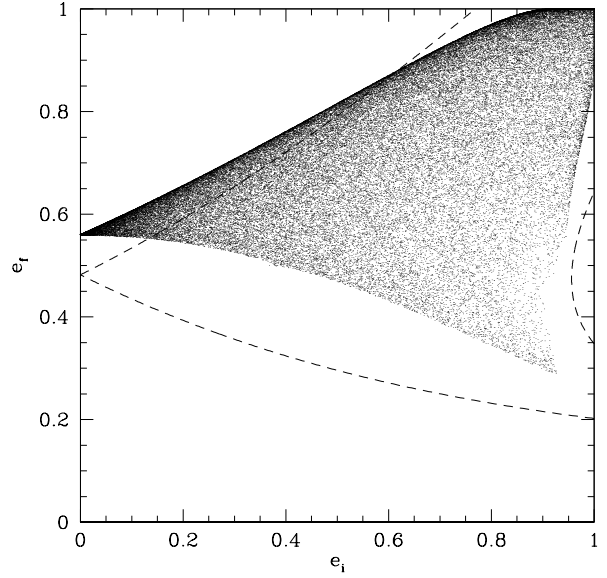}
\caption[MM.ps]{
The dashed lines enclose the range of possible final eccentricities, depending on initial eccentricity and longitude of
periastron, calculated in the limit of low eccentricity. The points show the solution using the full expression for $J$, which preserves the
basic structure of the low e expansion, but shifts the solution to higher values. These solutions are for the case
$e_b=0.1$, and will shift according to this quantity and the strength of the tidal dissipation.
\label{efi}}
\end{figure}

The size of the eccentricity jump is large enough, however, that we should use the full expression for $J_f$.
Figure~\ref{efi} shows the resulting range of final eccentricities as a function of initial eccentricity, for our nominal $Q_e' \sim 10$,
assuming $e_b=0.1$. We can see that, for small
initial eccentricities, there is a unique final eccentricity $\sim 0.56$, although this will, of course, be damped further by
tidal evolution after the resonance has been crossed. In the absence of further eccentricity excitation, this also predicts a 
final semi-major axis $\sim a_0 (1-e_f^2) \sim 0.0205$AU, for $a_0 = 0.03$~AU. The size of the eccentricity jump also scales
with the eccentricity of 55~Canc~b. If $e_b$ is as small as 0.01, then the final eccentricity is $\sim 0.18$ in the limit of
low initial eccentricity. Nevertheless, we see that substantial eccentricities can be generated even if the eccentricity of the
perturbing planets are relatively small. The eccentricity of Planet~b may also experience moderate pumping due to the coupling
between planets e, b and c via a third mode (see Figure~\ref{afg}), but this is limited by the large mass ratio.

We find a similar inclination excitation,
\begin{equation}
Z_f = Z_i + \frac{\pi \delta^2}{2 \left|\dot{h_0}\right|} + \delta \sqrt{\frac{2 \pi Z_i}{\left|\dot{h_0}\right|}} \cos \Omega_i.
\end{equation}
Since $\tau_0$ once again depends on the eccentricity, we note that the inclination resonance lies slightly outside the eccentricity
resonance, so that we expect this to scale with $e_i$, not $e_f$.
In the limit of low eccentricity and inclination, the analogue to equation~(\ref{ejump}) is 
\begin{equation}
i_f^2 = i_i^2 + \frac{\pi \delta^2 \tau_0}{2 h_0 \sqrt{a} e_i^2} + \delta \sqrt{\frac{2 \pi \tau_0}{h_0 \sqrt{a}}} \frac{i_i}{e_i} \cos \Omega_i.
\end{equation}
 In the limit of low initial eccentricity and inclination, this still
predicts a final inclination $\sim 45^{\circ}$, for values of 55~Canc~b of $e_b=0.1$ and $i_b=2^{\circ}$.

Therefore, this analysis supports the idea, expressed in the previous section, that substantial changes in inclination and eccentricity are possible
for reasonable planetary parameters. It also demonstrates the dependance on parameters, in that eccentricity and inclination changes increase with the
dissipation timescale $\tau_0$, i.e. crossing the resonance more slowly increases the excitation.

\subsection{Additional Planets}
\label{More}

The present analysis is based on the secular behaviour of the observed planets. However, the secular oscillations
depend on the effects of all the planets in the system, which raises the question whether it is possible to
substantially alter the properties of the system evolution if there are as-yet-undetected planets present.

\begin{figure}
\includegraphics[width=84mm]{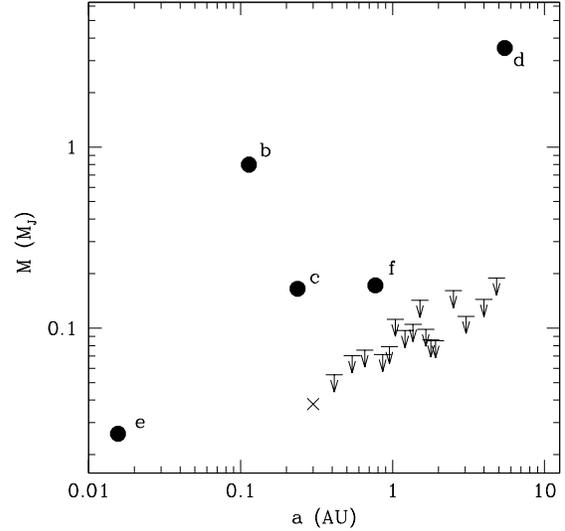}
\caption[Zlim.ps]{The solid points show the currently known planets of the $\rho$~55~Cancri system, with the
radial velocity values for the masses shown. The upper limits indicate the 90\% confidence mass limits for selected
regions of period space. Planets with this period and masses larger than this limit should have already been
detected in the radial velocity data. The cross shows the approximate location of the tentative 60.76~day signal
discussed in passing by Nelson et al. (2014).
\label{JonFig}}
\end{figure}

To assess this effect, we have taken the collection of extant radial velocity data from Nelson et al. (2014), Endl et al. (2012) and Fischer et al. (2008)
and determined how large a planet could remain undiscovered in this data. The orbital periods of the five
known planets were held fixed at the values determined by Nelson et al. but other parameters such as mass, eccentricity
and longitude of periastron were allowed to vary, and then the data was refit with a series of models with an additional planet
with a range of mass and semi-major axis. Orbital periods from 94 to 3900 days were searched, and the mass constraints were
defined as those that would yield a signature larger than a 90\% confidence level fluctuation with respect to the best fit.
The resulting mass limits are shown in Figure~\ref{JonFig}.

For planets in the range 0.4--0.75~AU, radial velocity signals would be detected for planets with
masses $> 4 \times 10^{-5} M_{\odot}$. If we include a single planet of this mass and period in
our secular evolution, it results in no qualitative changes to the evolution. This is because the extra
planet only exhibits a substantial coupling to two modes. The first is a mode dominated by the new
planet alone, and the second mode is the mode coupling all of the inner planets together (the
mode marked e-b-c in Figure~\ref{afg}). Thus, another planet in this gap would share in the mode
that mixes all the planets together, and can contribute to the angular momentum budget of the
secularly coupled system, but it does not substantially alter the structure of the modes
that undergo the $g_{ce}$ and $f_{ce}$ resonant interactions). We have tested
the effects of planets an order of magnitude larger than our RV limit and still find no substantial
qualitative change in the mode behaviour. Nelson et al (2014) note the presence of a weak signal
in their data with period 60.76~days and velocity semi-amplitude $\sim 2 m/s$. This is of similar
magnitude to the limits we find, and so, if such a planet exists, it would not substantially
change the secular structure of the system.

We have also tested the effects of a planet in the gap between planets~f and d. If such a planet
improves the secular coupling of the inner planets and the outer planets, it might change the
evolution of the system by providing a greater reservoir to store angular momentum and might
help to drive an even stronger evolution. However, with mass limits $<0.1 M_J$, extra planets
with $a > 1$~AU appear to only weaken any residual coupling between inner and outer planets.
New planets interior to this would be within the 3:2 commensurability with planet~f and should
provide strong dynamical signatures that would be detected in the radial velocities.

We therefore conclude that the behaviour discussed here is likely to be robust with respect
to the presence of as-yet-undetected planets in the 55~Cancri system.

\section{Numerical Exploration}
\label{Num}


The analytic and semi-analytic results offer a qualitative explanation for the
possible origin of a large inclination difference in the 55~Cancri system, but
there are several reasons to be cautious about the quantitative results. 
Our initial estimates suggest that eccentricities and inclinations can reach substantial values,
exceeding the approximations
 of classical secular theory. Furthermore, the
proximity of planets 55~Canc~b and 55~Canc~c to the 3:1 mean motion resonance
could possibly amend the secular oscillations and shift the resonance locations.
This can be incorporated into the classical formalism in the case of first order resonances (Malhotra et al. 1989)
but higher order resonances scale with higher powers of eccentricity and inclination, which 
would introduce nonlinearities into the classical secular equations.
Thus, we have performed numerical integrations of model systems with the Mercury6
code (Chambers 1999) to examine the secular structure of actual 55~Cancri analogues.

The standard Mercury6 algorithm does not account for relativistic precession nor
tidal evolution of the orbit. We have thus included a central potential contribution to
produce the former and account for the latter by instituting a drag force by including
a radial velocity damping term. The small stepsizes required to resolve these short period
orbits means that we are able to reproduce the expected precession rate without needing
to incorporate invididual stepsizes (e.g. Saha \& Tremaine 1994).
 We describe the calibration of the latter in appendix~\ref{NumCal}.

\subsection{Test Particle Orbits}

The simplest test of the effect of the 3:1 resonance is to examine the secular
oscillations of test particles interior to 55~Canc~b. We simulated several analogues
of the 55~Canc planetary system with 55~Canc~e replaced by a set of test particles
on co-planar circular orbits and with a range of semi-major axes, in order to understand the
secular forcing this planet would experience from the other four. We chose the 
longitudes of periastron and ascending node randomly for the four planets and integrated
several cases, to understand what sort of secular behaviours might be expected.

\begin{figure}
\includegraphics[width=84mm]{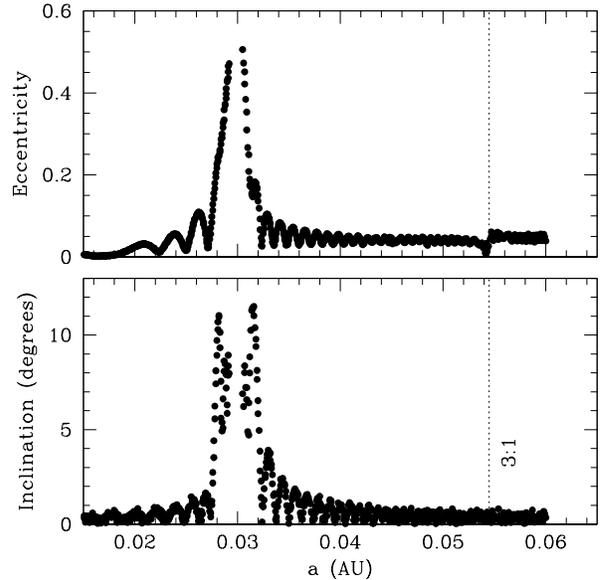}
\caption[Merc21_test.ps]{
The upper panel shows the eccentricity, while the lower
panel shows the inclination, after 1.1~Myr of evolution. Further evolution can drive
the eccentricity to high enough values to impact the star.
\label{Merc21_test}}
\end{figure}

Figure~\ref{Merc21_test} shows the expected result, based on the considerations of
\S~\ref{Tides}. We start with 800 test particles on circular, coplanar orbits between
0.015 and 0.065~AU, perturbed by the four planets of the 55~Cancri system with a$>0.1$~AU.
The figure shows the resulting forced eccentricity and inclination after 1.1~Myr of
evolution. Over most of this range, the appearance of simple Lagrange oscillations are
evident, but at $a \sim 0.03$AU, both eccentricity and inclination grow due to the
resonance of the free Lagrangian precession of the test particles with the secular oscillations
of the forcing planets.

\begin{figure}
\includegraphics[width=84mm]{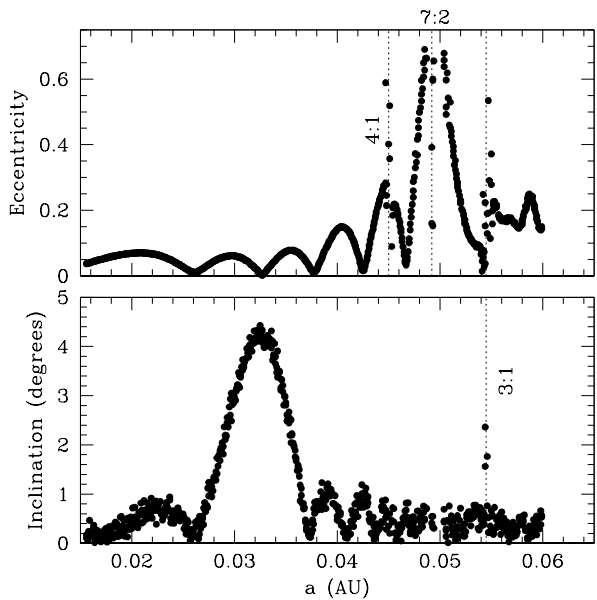}
\caption[Merc22_test.ps]{The upper panel shows the eccentricity, while the lower
panel shows the inclination, after 0.3~Myr of evolution. In this case we see that
perturbations due to the proximity of the 3:1 resonance have shifted the location
of the eccentricity resonance without shifting the location of the inclination
resonance markedly. The larger eccentricities also lead to enhanced contributions
from mean motion resonances with 55~Canc~b as well.
\label{Merc22_test}}
\end{figure}

However, other behaviour is also possible. Figure~\ref{Merc22_test} shows the evolution
of the same test particle setup but forced by a planetary system with a different choice
of eccentricities and longitudes of periastron and ascending node. In this case the 
inclination resonance occurs at approximately the same location, but the eccentricity
resonance occurs at $a \sim 0.05$AU. Further examination of the results from this integration
shows that the resonant angles generated by the pair 55~Canc~b and 55~Canc~c circulate most
of the time, but undergo intermittent passages through resonance, which occur when the secular
oscillations bring one of the planetary eccentricities close to zero.
 The origin of this behaviour is discussed in more detail
in appendix~\ref{Almost}, but the salient feature for this analysis is that this interaction
between resonant and secular behaviour is enough to shift the secular resonance in some cases.

\subsection{Tidal Evolution}
\label{NumTides}

In order to examine the consequences of this kind of evolution for plausible
55~Cancri origins,  we perform
the following numerical experiment on the 55~Cancri system, using the Mercury6 code.

We integrated 100 systems with 55~Canc~e originally located at 0.033~AU (i.e.
just outside the secular resonance location in the classical estimate). The other four planets assume the
semi-major axes determined by Nelson et al. (2014) from the latest compilation of the data. We assume
masses for the four outer planets assuming an inclination angle of 53$^{\circ}$ (i.e. an
enhancement by a factor of 1.25 above the nominal edge-on estimate), although we retain the nominal mass
of 55~Canc~e because it is transiting and therefore has inclination of 90$^{\circ}$. Tidal dissipation
tends to align the longitudes of pericenter, 
 so we draw the initial 
 longitudes of perihelion and ascending node, as well as the
mean longitude at the start, from uniform distributions. The inclinations are chosen from a
Gaussian distribution with dispersion of $2^{\circ}$, which is consistent with estimates for the flatness
of compact planetary systems based on Kepler studies (Lissauer et al. 2011a; Tremaine \& Dong 2012; Fang
\& Margot 2012). Tremaine \& Dong (2012) note that larger inclination dispersions are possible, but our
assumption represents a conservative starting point in order to illuminate the size of the effect produced
by the secular resonance. Although the 55~Cancri planets have measured eccentricities, these could have
been substantially reduced by the effects of tidal damping and secular coupling. In order to assess the
strength of the possible interactions, we draw our initial eccentricity distribution from the eccentricities
observed in other planetary systems. Several proposals have been made for describing the exoplanet eccentricity
distribution (e.g. Shen \& Turner 2008) but including planets at large orbital periods and single planets may 
overestimate the eccentricities for planets in compact multiple systems. Wang \& Ford (2011) estimate the eccentricities for single planets with
short orbital periods but do not include multiples. Limbach \& Turner (2014), on the other hand, derive
the eccentricity distribution as a function of multiplicity, but include systems on large scales. We
opt to draw 
 our eccentricities from the `short period' beta distribution fit by 
Kipping (2013) to the observed exoplanet data with orbital periods $<382$~days. This draws values from the cumulative probability
\begin{equation}
P(e) = \frac{\Gamma (3.967)}{\Gamma (0.697)\Gamma (3.27)} e^{-0.303} \left( 1 - e \right)^{2.27}.
\end{equation}
We adopt this as representative  of the
potential initial conditions in the 55~Cancri system as it includes the bulk of the 55~Cancri planets
(except for 55~Canc~d) within its period ($< 387$~days) threshold. The systems are then integrated for 1~Myr.
Our implementation of tidal evolution in Mercury6 uses the force law used by Hut (1981),
to damp eccentricity. As discussed in appendix~\ref{NumCal} we calibrate the dissipation in order to produce an eccentricity
damping timescale of $1.3 \times 10^5$~years for an 8$M_{\oplus}$ planet at a=0.033~AU. For a planetary radius of $R=2 R_{\oplus}$,
this is equivalent to $Q'=13$.

\begin{figure}
\includegraphics[width=84mm]{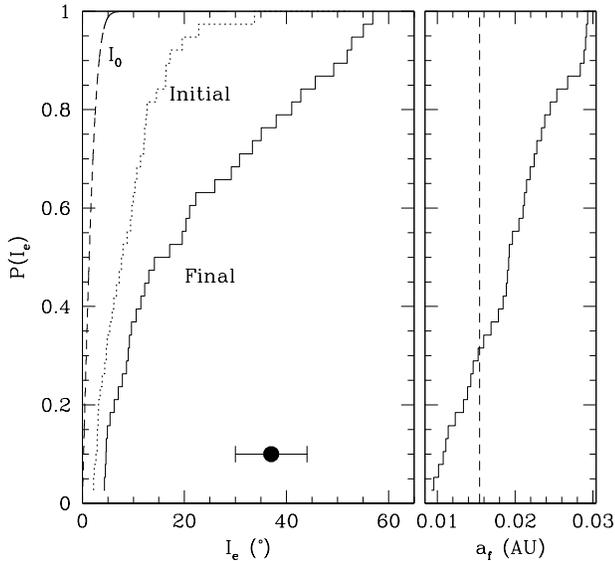}
\caption[Incs.ps]{The right-hand panel shows the distribution of semi-major axis for the 38 systems whose evolution
puts them interior to 0.03~AU. The observed location of 55~Canc~e is shown as a vertical dashed line. The left-hand
panel shows the final inclinations of these same systems (solid histogram). The dashed histogram is the distribution
from which the initial planetary osculating inclinations were drawn. Because of secular forcing, the actual initial
proper inclinations are larger, and are shown by the dotted histogram.
 The solid point with error bars indicates the inclination of 55~Canc~d measured
by McArthur et al. (2004), relative to an edge-on orbit, such as that of 55~Canc~e.
\label{Incs}}
\end{figure}

Of the 100 planetary systems integrated, 33 underwent dynamical instability, usually after less than 0.1~Myr
(where we define this as undergoing planetary scattering and collisions which removed at least one and often
several planets from their observed orbits -- be it by ejection or collision with the star or another planet).
This indicates that even the reduced eccentricities of Kipping's `short period' distribution are sometimes too large
for a system as closely packed as the 55~Cancri system. Of the 67 systems whose planets survive to the present,
38 reach semi-major axes $< 0.03$~AU within 1~Myr, while 29 do not. Those that pass within 0.03~AU usually
lie well inside 0.03~AU, as shown in the right-hand panel of Figure~\ref{Incs}. This is a consequence of the
eccentricity jump due to the secular resonance and the subsequent circularisation of the orbit. Figure~\ref{Incs}
also shows the resulting inclination distribution of the 38 systems with final semi-major axis $<0.03$AU. We
find inclinations can be pumped up to almost 60$^{\circ}$, in agreement with the simple estimates of earlier
sections. It should be noted that the inclinations shown in Figure~\ref{Incs} are the amplitudes of the secular
oscillations, not the instantaneous values, which is why the initial conditions (dotted histogram) are larger than the seed distribution
(dashed histogram) -- the secular forcing from the other planets drives the inclination of 55~Canc~e up even
if the initial value is small.

\begin{figure}
\includegraphics[width=84mm]{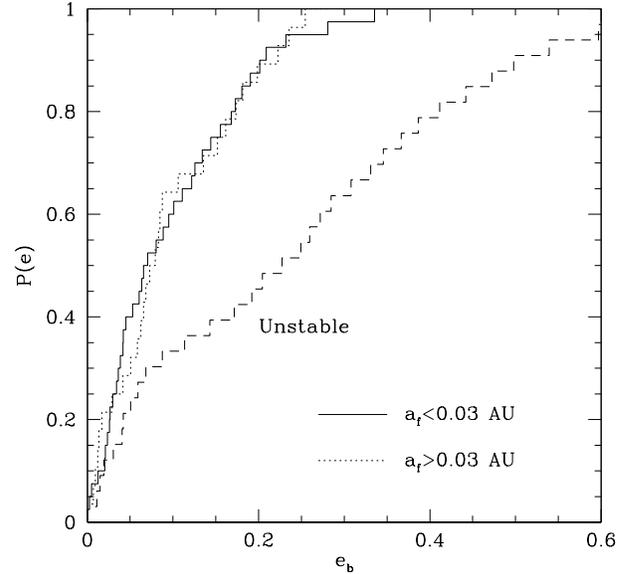}
\caption[Edis.ps]{The solid histogram shows the initial eccentricity of planet 55~Canc~b in the case where 
the planet e evolves to semi-major axis $<0.03$AU within 1~Myr. The dotted histogram shows the initial
eccentricity in the case where the system remains stable but does not satisfy this criterion. The dashed
histogram shows the initial conditions that lead to dynamical instability and shows that eccentricities
$>0.25$ for planet~b usually lead to instability. The similarity of the dotted and solid histograms suggest
that it is not simply a question of secular amplitudes that determines the tidal evolution of the system.
These distributions are consistent with a Gaussian distribution of eccentricities with dispersion = 0.12. 
\label{Edis}}
\end{figure}

Of the 29 stable systems that do not penetrate inside 0.03~AU within 1~Myr, it is possible that some may just
have insufficient amplitude in the relevant modes to secularly pump the eccentricity of 55~Canc~e to values that
promote tidal evolution. However, the distribution of initial eccentricities for 55~Canc~b, which is the principal
driver of the secular pumping, is quite similar to those cases which do migrate, as shown in Figure~\ref{Edis}. This
suggests that many of the systems have sufficient amplitude, but have had their eigenvalues shifted enough
by the 3:1 resonance that it lies exterior to the starting location of our simulation.

To test this hypothesis, we repeat the integration of 10 of the systems which do not undergo sufficient evolution,
with all of the same parameters except that the initial semi-major axis of 55~Canc~e is taken to be 0.045~AU
instead. This is to place it closer to the region destabilized in the case shown in Figure~\ref{Merc22_test}.
Of these, two exhibited little evolution, but in another five, 55~Canc~e experienced such strong eccentricity excitation that
it collided with the star with little semi-major axis evolution or inclination excitation. In the remaining three,
the inner planet achieved a 55~Canc~e analogue orbit, with a circularised orbit with $a<0.03$~AU, and maximum 
inclinations of 8.2, 12.7 and 18.2$^{\circ}$. Thus, in many of these systems, the inner planet experiences rapid
eccentricity growth, to the point that it rises to a catastrophic level before tides even play a role. This is
likely due to the additional forcing discussed in appendix~\ref{Almost} due to the 3:1 resonance between 55~Canc~b and c.

\begin{figure}
\includegraphics[width=84mm]{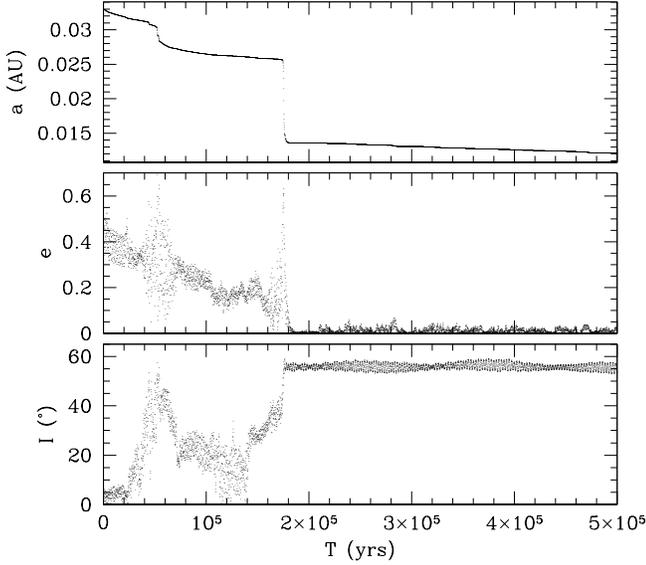}
\caption[Example.ps]{The upper panel shows the evolution of the semi-major axis, the middle panel the evolution
of the eccentricity, and the lower panel the evolution of the inclination. The case shown here proceeds along the
normal evolutionary path, with both eccentricity and inclination resonances occurring near 0.03~AU.
\label{Example}}
\end{figure}

\begin{figure}
\includegraphics[width=84mm]{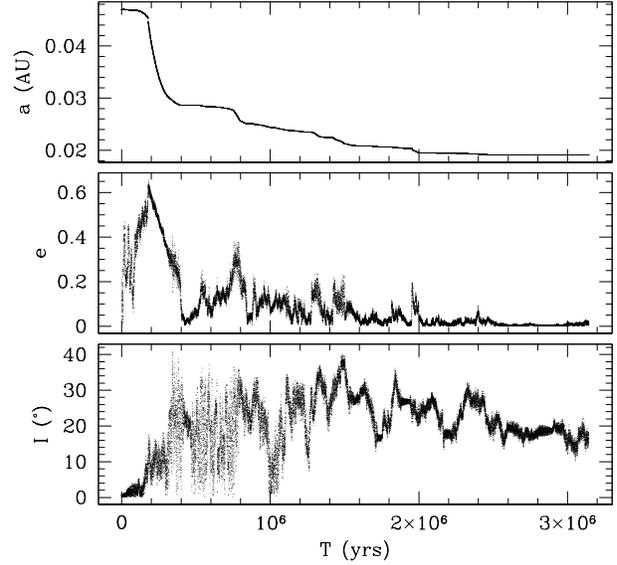}
\caption[Example2.ps]{The upper panel shows the evolution of the semi-major axis, the middle panel the evolution
of the eccentricity, and the lower panel the evolution of the inclination. In this case, the eccentricity resonance
is shifted outwards, to $\sim 0.043$~AU.
\label{Example2}}
\end{figure}

Figures~\ref{Example} and \ref{Example2} show two examples of the kind of evolution that might produce a 
misaligned 55~Cancri~e. In Figure~\ref{Example} we show an evolution from 0.033~AU which follows the path
predicted by classical secular theory, in which a steady inward migration is punctuated with sudden jumps
in eccentricity and inclination substantially in excess of the normal secular oscillations, resulting in
a final inclination of almost $60^{\circ}$ with respect to the original orbital plane. Figure~\ref{Example2} shows
the evolution in the case where the eccentricity is pumped up to large values even at a starting location of
0.045~AU, which then drives tidal migration and again excites inclination as it passes through the unshifted
inclination resonance.

\subsection{Effect of the Companion}

The presence of an M-dwarf companion at $>1000$~AU can, in principle, induce precession of the
planetary system (Innanen et al. 1997; Kaib et al. 2011; Boue' \& Fabrycky 2014) if the orbit
is sufficiently eccentric and inclined. However, the precession timescales are long compared to
the time it takes 55~Canc~e to cross the resonance, and so the orbit tilting discussed here is 
insensitive to the influence of the external companion. We have verified this by repeating five
of the simulations which crossed the resonance but including a 0.26$M_{\odot}$ companion with
semi-major axis 1250~AU, eccentricity$=0.93$ and inclination $115^{\circ}$ as discussed by 
Kaib et al. (2011). In all cases the eccentricity and inclination excitation  was reproduced, with rapid eccentricity pumping
followed by circulation, and the final orbital tilt of 55~Canc~e relative to the other planets.

\section{Results}

Our principal goal in this paper is to examine the effect of secular interactions coupled
with tidal evolution on the configuration of the 55~Cancri planetary system. The generic
behaviour of multi-planet systems under the influence of tidal evolution is that the
eccentricities of several planets can be reduced, while the bulk of the semi-major axis
evolution is borne by the innermost planet (Wu \& Goldreich 2002; Greenberg \& van Laerhoven 2011;
Hansen \& Murray 2014). Such behaviour in this system would imply that 55~Canc~e began with a larger
semi-major axis and migrated to the one observed today.

If 55~Canc~e began with a semi-major axis $>0.033$AU, it will have crossed multiple secular
resonances, dominated by the pair of planets 55~Canc~b and 55~Canc~c, which could have excited
both substantial eccentricities and inclinations. We have demonstrated these effects using
both classical secular theory, analytic estimates and direct numerical integrations. These
suggest that inclinations of $30-40^{\circ}$ between 55~Canc~e and the other planets in
the system are quite possible for initial conditions that draw planetary eccentricities
from a Gaussian with dispersion $\sigma_e = 0.12$ (an approximate fit to the dynamically
stable portion of our simulations -- see Figure~\ref{Edis})
 and mutual inclinations from a Gaussian distribution with a $\sigma_i = 2^{\circ}$ dispersion
(consistent with the flat planetary systems observed by the Kepler satellite). 
The substantial amplification is a consequence of the large mass ratio between 55~Canc~e
and 55~Canc~b, which causes any coupling between the planets to be reflected disproportionately in the
smaller of the pair. The median eccentricity of the input planets, once the dynamically unstable
systems are removed, is $\sim 0.07$, which is similar to the value expected for a five
planet system ($\sim 0.085$) from the power law
fit of Limbach \& Turner (2014). 

The ability to generate a substantial inclination offset from the orbital plane of the original
planetary system offers an explanation for the tension between the astrometric measurement
of 55~Canc~d's inclination by McArthur et al. (2004) and the transitting nature of 55~Canc~e (Winn et al. 2011;
Demory et al. 2011). One observation potentially in conflict with this scenario is the statistically marginal detection
of an extended H atmosphere of 55~Canc~b by Ehrenreich et al. (2012). If confirmed, this would suggest a different
configuration for the system, with an approximate alignment of 55~Canc~e and b.

\subsection{Planetary Heating}

The passage through the eccentricity resonance causes a rapid increase in the eccentricity of 55~Canc~e and
a corresponding rapid decrease in the periastron distance. The strong radial dependance of the tidal force means
that this leads to a sharp increase in the rate of tidal dissipation and tidal heating, which in turn may
substantially influence the structure of the planet. In particular, we note that the amount of energy available
could potentially have removed a substantial amount of Hydrogen from 55~Canc~e.

 To estimate the impact
of the heating on the structure of the putative planet, we note that the final planetary semi-major axis is
roughly half that of the secular resonance, so that the binding energy dissipated as a result of the resonance
crossing is $>10^{42}$ ergs. This is comparable to the gravitational binding energy of a Neptune mass planet,
and so can substantially influence the structure of the planet.

The 55~Cancri system already contains several planets, such as 55~Canc~c and 55~Canc~f, which have masses similar
to Neptune or Uranus. Thus, 55~Canc~e may potentially represent the rocky core of a thermally altered version of
this class of planet. To ascertain the likelihood of this, we consider
 initial
parameters for 55~Canc~e, of a mass of 20$M_{\oplus}$ and a radius of $4 R_{\oplus}$.
 The rate at which this energy is dissipated will depend on the
rate of tidal dissipation, which we take to be characterised by $Q' \sim 10^4$, as in the case of Neptune (Banfield \& Murray 1992;
Zhang \& Hamilton 2008)
Assuming an average eccentricity $\sim 0.5$ during this period, and the constant time-lag model (Hut 1981)
for tidal dissipation, we find that the binding energy is dissipated over a period $\sim 10^6$~years, yielding
an average luminosity $\sim 10^{-4} L_{\odot}$.

The effective temperature of a $4 R_{\oplus}$ planet, radiating at $10^{-4} L_{\odot}$, is $\sim 3000$K, comparable
to the hottest estimated equilibrium temperatures of irradiated giant planets (e.g. Konacki et al. 2003; Hebb et al. 2009; Gillon et al. 2014), and $\sim 40\%$ larger than the equilibrium
temperature maintained by irradiation for this planet. Nevertheless, as in the case of  hot Jupiters on these scales,
mass loss by Jeans escape is still not sufficient to substantially alter the planet mass, because
$\lambda \sim G M m_p/R k T \sim 12$ for such a planet with this effective temperature.

However, dissipation does not only increase the temperature of the planet.
We have previously reviewed the impact of strong tidal heating on gaseous giant planets (Hansen 2012) and found that the
principal repository of the energy at high levels of dissipation is in the binding energy of the planet  (i.e. at the
 level of dissipation relevant here, the radius of the planet increases, and central temperature can even drop).
 For the
parameters of relevance here, the radius need only expand to $\sim 6.5 R_{\oplus}$ to begin Roche lobe overflow
at periastron
and initiate mass loss.

\begin{figure}
\includegraphics[width=84mm]{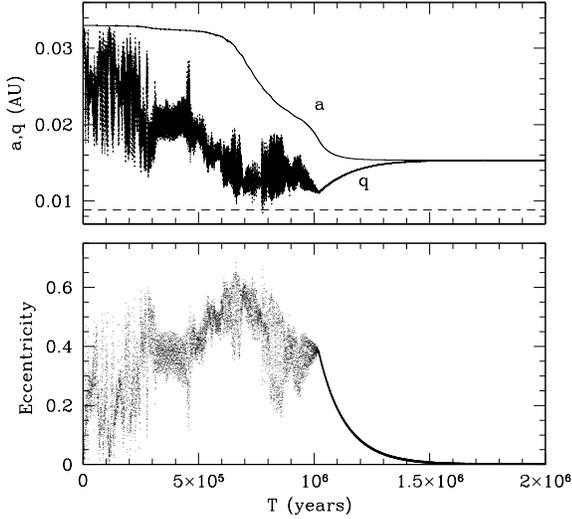}
\caption[MercNep1.ps]{The upper panel shows the evolution of both the semi-major axis (upper, smooth, curve) and the
periastron (lower curve) for a $20 M_{\oplus}$ planet driven to high eccentricity by the crossing of a secular resonance.
The horizontal dashed line indicates the semi-major axis at which a $20 M_{\oplus}$, $4 R_{\oplus}$ planet will fill its
Roche lobe. The lower panel shows the corresponding eccentricity excitation.
\label{MercNep1}}
\end{figure}

 To demonstrate this, we have repeated one of our simulations but replaced 55~Canc~e with a 20$M_{\oplus}$ object, and
adjusted the level of dissipation to mimic a value of $Q' \sim 10^4$ as appropriate for Neptune. Assuming that the 
radius of the planet is now $4 R_{\oplus}$, this requires weakening the numerical dissipation by only a factor of 60,
because $Q'$ scales with the fifth power of the planetary radius. The resulting evolution is shown in Figure~\ref{MercNep1}.
We see that the character of the evolution is the same, with excitation to high value of eccentricity and a sharp
drop in semi-major axis, although a little less steep, as befits the weaker dissipation. We also show the evolution of the
periastron during this period and note that it skirts the edge of Roche-lobe overflow even for an un-inflated radius of
$4 R_{\oplus}$. Thus, even a small amount of radius inflation due to the tidal dissipation will drive this planet into a
mass loss event.

Thus, it is possible that the present mass and radius of 55~Canc~e are the result of mass loss, probably through Roche lobe
overflow, induced by the strong tidal heating generated by the passage through the secular eccentricity resonance.
If this is the case, 55~Canc~e may originally have been a Neptune-mass planet, more similar to other known members of
the system, such as 55~Canc~c and 55~Canc~f, suggesting migration as a chain of giant planets and
failed cores. Residual heating may still make a contribution to the amplitude of
the secondary eclipse (Bolmont et al. 2013) if the eccentricity is in the range $\sim 0.001-0.01$.

\subsection{Other Systems}

The results above suggest that the 55~Cancri system may exhibit a substantial inclination dispersion, but this
is also a relatively unusual planetary system, as most short period Jupiters do not reside in systems of non-hierachical high
multiplicity (Wright et al. 2009; Steffen et al. 2012). Thus, it is of interest to enquire as to the frequency of such secular resonances
in compact planetary systems.
Table 2 shows examples of several systems which may possess
such a resonance.

For systems detected via radial velocities, we examine those systems which possess three or more planets 
inside 0.5~AU, at least two of which lie outside 0.05~AU. In addition to 55~Cancri, this yields 61~Vir (Vogt et al. 2010), HD~40307 (Mayor et al. 2009) and HD~10180 (Lovis et al. 2011) around solar-type stars, and
 GJ~876 (Rivera et al. 2010) around an M star. In each case, we consider the secular architecture of the system given the
nominal semi-major axes and masses, and examine how this changes as we vary the semi-major axis of the innermost
planet. 
 This accounts for possible motion due to tidal evolution, which is reflected primarily in the evolution
of the innermost planet for secularly coupled systems (Hansen \& Murray 2014). However, some care is necessary as several of these systems contain uncertain detections. For HD~40307, we have adopted the more 
conservative configuration because the more uncertain planets (Tuomi et al. 2013) reside further out and including them turns out
to not significantly change the results. In the case of HD~10180, we include the uncertain innermost planet as a potential analogue to 55~Canc~e. We note also that more strongly resonant systems, like GJ~876, may exhibit more complicated
behaviour due to untreated resonant effects.

With these caveats, we find that each of the examined systems possess
at least one secular resonance in both inclination and eccentricity in the region through which a planet would
pass due to tidal evolution. In some cases, such as the 61~Vir and HD~10180 systems, the inner planet
would appear to actually be near the location of a secular resonance at present, offering a potential explanation
for non-zero eccentricities despite proximity to the star yielding strong tidal effects.

\begin{figure}
\includegraphics[width=84mm]{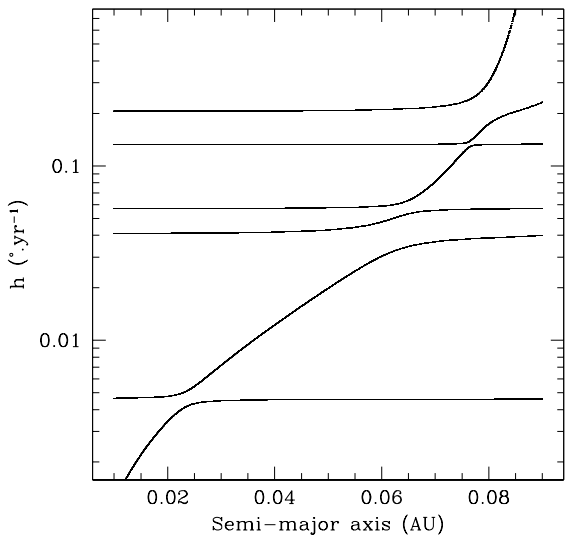}\caption[Kep11.ps]{The curves show the change in the secular inclination
eigenfrequencies
of a hypothetical planetary system constructed by adding an interior $1 M_{\oplus}$ planet to the known planetary
system Kepler-11. As the semi-major axis of the additional planet is varied, we see that there is an avoided
crossing at $\sim 0.065$~AU.
\label{Kep11}}
\end{figure}

\begin{table*}
\centering
\begin{minipage}{140mm}
\caption{Orbital periods of possible secular resonances in exoplanetary systems,
where P$_e$ indicates eccentricity resonances and P$_i$ indicates inclination
resonances. Numbers
in parentheses are caused by first order mean motion resonances. The final column
for the RV planets shows the observed orbital period of the currently known exoplanet.
 \label{SecTab}}
\begin{tabular}{@{}lclclc@{}}
System & P$_e$ & P$_i$ & P\\
 & (days) & (days) & (days) \\
\hline
RV Systems, assuming inner planet migration \\
\hline
$\rho$~55~Cancri & 1.66 & 2.0 & 0.74  \\
GJ~876 & 7.1, (14.7) & 7.3, 9.8,  17.0 & 1.9 \\
61~Vir & 4.2, (18.5) & 6.8 & 4.2\\
HD~40307b & 0.9, (4.7) & 2.2, 3.9 & 4.2\\
HD~10180 & 1.2, (3.0) & 0.8, 1.6 & 1.2\\
\hline
Transitting Systems, assuming an undetected $1 M_{\oplus}$ inner planet \\
\hline
KOI-94 & 0.9, (1.87) \& 2.1 & 1.2 \& 2.1\\
Kepler-11 & 2.4, (5.1), 8.0 & 1.3, 6.1, 8.0 \\
Kepler-30 & (13.9), 14.8 & 14.8\\
Kepler-9 & 1.0, 2.5 & 7.5 \\
Kepler-51 & $\cdots$ & $\cdots$ \\
\hline
\end{tabular}
\end{minipage}
\end{table*}

In the case of systems detected via transits, passage through an inclination resonance is likely to prevent
a planet from being observed in transit, assuming the natal planetary system is edge-on 
in the first place. Such excitations may explain inner holes in otherwise densely packed systems observed
by Kepler. An example is the Kepler-11 system (Lissauer et al. 2011b), which shows 6 planets, five of which are between 0.09--0.3~AU,
but none interior to 0.09~AU. If we examine the secular architecture of this system assuming the presence of
a seventh, $1 M_{\oplus}$ planet located in the interior hole\footnote{Such a planet would yield transit timing
variations of amplitude $< 1$ minute for Kepler-11b (Agol et al. 2005), well below the current timing
accuracy (Lissauer et al. 2013), as long as it lies interior to the 2:1 mean motion resonance.}, we find avoided crossings in both eccentricity
and inclination near semi-major axes $\sim 0.065$~AU, and other secular inclination resonances at
0.0233~AU and 0.0763~AU. Thus, it is possible that Kepler-11 contains
a low mass planet in the observed inner 'hole', but which is inclined relative to the outer planets. The inclination
could be substantial if the current period is $< 8$~days. The Kepler planet sample contains a large number of multiple
planet systems with potential for secular resonances. KOI-94 (Weiss et al. 2013) is interesting because of the presence of a Saturn mass planet in a similar
orbit to 55~Cancri. The innermost planet lies just outside a pair of secular resonances at 3.0~days (eccentricity) and
3.4~days (inclination) and so is consistent with a lack of inclination excitation. If we allow for the existence of a 1$M_{\oplus}$ planet interior to this, eccentricity resonances
are found at 0.9 and 2.1 days (and a mean motion resonance at 1.9~days) and inclination resonances at 1.2 and 2.1~days.
Thus, a planet that has tidally migrated to orbital periods $<$2~days in this system could have had its inclination
pumped up and have avoided transit. Another compact multiplanet system containing a massive planet is Kepler-30 (Fabrycky et al. 2012). However, the only resonances encountered by a tidally migrating $1 M_{\oplus}$ inner planet lie
at 14.8 days (both inclination and eccentricity) and at 13.9~days (a mean motion resonance). These may be too far
out to be strongly affected by tides. Kepler-9 (Holman et al. 2010), shows two Saturn-class planets and an interior
low mass planet, that we will assume to be $1 M_{\oplus}$ again. Since this inner planet does indeed transit, we must
assume little inclination excitation has taken place in this system. Indeed, the sole inclination resonance for this
configuration lies at 7.5~days, so that the planet would not experience any excitation if it started the migration
interior to this point. Eccentricity resonances are found at 1.0 and 2.5~days, bracketing the observed period.
Kepler-51 (Steffen et al. 2013) is an example of a compact Kepler planetary system with a large inner hole but no
Jovian-class planets. In this case, an inner $1 M_{\oplus}$ planet experiences no secular resonances. There
is a frequency commensurability at 2.9~days, but no avoided crossing because the modes are almost completely
decoupled (one is driven primarily by the relativistic precession). 

These results are necessarily incomplete because our knowledge of the planetary systems could also be
incomplete, but they illustrate the potential impact of secular resonances on the internal planetary configurations of multi-planet systems. They may be less important for lower mass systems where the precessions are weaker and relativistic effects can dominate at larger distances, but could generate substantial inclinations in the cases where Jovian-class planets are present on small scales.  Alternatively, the effect of stellar spin-down from an initially rapid  rotation may, in some cases, also induce secular
resonant crossings, as was once proposed for the eccentricity and inclination of Mercury (Ward, Colombo \& Franklin 1976).

\section{Conclusions}

We find that it is possible to explain the conflicting observations that 55~Canc~e transits, while
55~Canc~d is inclined by $\sim 53^{\circ}$, in terms of tidal decay of the 55~Canc~e orbit that causes
it to cross multiple secular resonances, which drive the planet to large eccentricities and inclinations.
van Laerhoven \& Greenberg (2012) have previously cast doubt on whether the system can have undergone
tidal evolution because not all of the secular modes coupled to the inner planet are completely damped.
However, a system need not have reached it's final state of a `fixed point' configuration, if
the evolutionary timescale is not too short. Indeed,
Hansen \& Murray (2014) found many model systems which were only partially damped, and the
large amplitude librations about secular alignment found by Nelson et al. are characteristic of this
stage of evolution. Furthermore, it worth
noting that the updated parameters of Nelson et al. do differ in important ways from the solutions
analysed by van Laerhoven \& Greenberg (2012). The pericenter alignment of the outer, decoupled, pair of
planets, which appeared to be coincidental, is no longer supported by the data. The remaining eccentricity
of 55~Canc~c suggests that this system is indeed still capable of further tidal evolution, although the
presence of a tentative signal (Nelson et al. 2014) between planets 55~Canc~c and 55~Canc~f may hint at additional influences
that can masquerade as an eccentricity.

The evolution through secular resonance increases the eccentricity as well as the inclination for this
planet, and can result in a substantial amount of heating within 55~Canc~e. This has potentially important
consequences for the use of 55~Canc~e as one of the highest signal-to-noise examples of a super-Earth planet.
We have shown that it is possible that 55~Canc~e could have lost mass through Roche lobe overflow if it expanded
due to tidal heating, and so 55~Canc~e may represent a class of disrupted planetary cores that have been
occasionally invoked  by several authors (Gu, Lin \& Bodenheimer 2003; Jackson, Barnes \& Greenberg 2009; Hansen 2012; 
Valsecchi, Rasio \& Steffen 2014), rather than a rocky planet
that assembled through planetesimal collisions. A more massive and volatile-rich initial mass would make 55~Canc~e
 comparable with
the other sub-jovian planets in the system, and suggests a scenario in which 55~Canc~b migrated inwards with a
retinue of Neptune-mass attendants, one of which was subsequently heated and stripped to the planetary core we
observe day.

We have also shown that many compact multiple planet systems exhibit possible secular resonances interior to
0.1~AU, suggesting that the phenomenon of moderate-to-highly inclined inner planets may be quite common. 
Such inclination pumping could potentially explain  the observed lower deficit of multiple transitting planets at very short periods in the
Kepler sample (Steffen \& Farr 2013). Similarly, pumping of eccentricity and subsequent tidal circulation is a potential
mechanism for moving planets to very short orbital periods and promoting their disintegration (Rappaport et al. 2012, 2014; 
Jackson et al. 2013).

The analysis presented here is predicated on the HST astrometric detection by McArthur et al. (2004), which places
some constraints but is based on a limited data span. Further astrometric observations of this system, either with the
HST Fine Guidance sensor or the GAIA spacecraft, would provide a welcome improvement on this constraint, and could
help to refine the parameters. A substantial mutual inclination between 55~Canc~e and the rest of the planets could also
be tested with transit monitoring, to search for precession of the orbit.

This research has made use of NASA's Astrophysics Data System and of the
the NASA Exoplanet Archive, which is operated by the California Institute of Technology, under contract with the
National Aeronautics and Space Administration under the Exoplanet Exploration Program. The authors
also acknowledge the contributions from a useful referee report.

\newpage

\appendix

\section{Eccentricity and Inclination Excitation due to secular resonance crossing}
\label{MM11}

We consider an extended version of the Hamiltonian discussed in Minton \& Malhotra (2011),
\begin{eqnarray}
H & = & H_e + H_i  \\
& = &  -g_0 J - h_0 Z + \epsilon \sqrt{2 J} \cos (\bar{\omega}_p - \bar{\omega})
 + \delta \sqrt{2 Z} \cos (\Omega_p - \Omega) \nonumber
\end{eqnarray}
where $J = \sqrt{a} (1 - \sqrt{1 - e^2})$,  $Z = \sqrt{a (1 - e^2)} (1 - \cos i)$, $g_0$ and $h_0$ are the precession frequencies for the longitude of periastron $\bar{\omega}$, and longitude of the ascending node $\Omega$ respectively. The parameters
$\epsilon$ and $\delta$ represent the strength of the coupling to the secular modes, whose orientations are represented by $\bar{\omega}_p$,
 and $\Omega_p$.

We express these quantities as
\begin{eqnarray}
 g_0 & = & n \left[\frac{1}{4}  \sum_j \frac{m_j}{m_c} \alpha_j^2 b_{3/2}^{(1)} (\alpha_j) + \frac{3 G m_c}{c^2 a} \right] \\
 h_0 & = &  \frac{n}{4} \sum_j \frac{m_j}{m_c} \alpha_j^2 b_{3/2}^{(1)} (\alpha_j) \\
 \epsilon & = & \frac{n}{4} \sqrt{G m_c a}  \sum_j \frac{m_j}{m_c} \alpha_j^2 b_{3/2}^{(2)} (\alpha_j) E_j \\
 \delta & = & \frac{n}{4} \sqrt{G m_c a}  \sum_j \frac{m_j}{m_c} \alpha_j^2 b_{3/2}^{(1)} (\alpha_j) I_j 
\end{eqnarray}
where $E_j$ and $I_j$ are the eigenvector contributions from each planet to the mode under discussion, and $g_0$ includes
the effect of relativistic precession to lowest order.

The approach of MM11 is to perform a pair of canonical transformations to variables x and y, represented
by $x = \sqrt{2 J} \cos \phi$ and $y = \sqrt{2 J} \sin \phi$, where $\phi = \bar{\omega}_p(t) - \bar{\omega}$.
In our case, it is $\bar{\omega}$, rather than $\bar{\omega}_p$, that is a function of time, but the formalism remains the same. However,
we must now incorporate changes in $\epsilon$ concommitant with the changes in $g_0$, since both are affected
by the radial migration of the planet.

Following the formalism of MM11, we find that the Hamiltonian equations of motion yield 
\begin{eqnarray}
 \dot{x} & = & \dot{g}_0\, y\, t \\
 \dot{y} & = & - \epsilon - (\dot{g}_0 x + \dot{\epsilon}) t
\end{eqnarray}
which is the same as equations~(9) and (10) of MM11, except for the $\dot{\epsilon}$ term. 

The method of solution is similar, whereby one can find a homogeneous solution in terms
of trigonometric functions of $\dot{g}_0 t^2$, and then substitute back to find a solution
to the inhomogeneous version by solving for the multiplicative constants as functions of
time. Thus, we assume
\begin{eqnarray}
 x(t) & = & A(t) \cos \dot{g}_0 t^2 + B(t) \sin \dot{g}_0 t^2 \\
 y(t) & = & - A(t) \sin \dot{g}_0 t^2 + B(t) \cos \dot{g}_0 t^2 
\end{eqnarray}

MM11 solved for A and B in terms of the Fresnel integrals
\begin{eqnarray}
 S (t) & = & \int_0^t \sin z^2 dz \\
 C (t) & = & \int_0^t \cos z^2 dz.
\end{eqnarray}
The presence of $\dot{\epsilon}$ term introduces a slight modification to the
solution, yielding
\begin{eqnarray}
 A(t) & = & \frac{\epsilon}{\sqrt{\left|\dot{g}_0\right|}} S(t') + \frac{\dot{\epsilon}}{\left|\dot{g}_0\right|} S'(t')  \\
 B(t) & = & -\frac{\epsilon}{\sqrt{\left|\dot{g}_0\right|}} C(t') - \frac{\dot{\epsilon}}{\left|\dot{g}_0\right|} C'(t') 
\end{eqnarray}
where $t' = \sqrt{\left|\dot{g}_0\right|} t$, and $S'$ and $C'$ are modified versions of the Fresnel integrals, such that
$$
S'(t) = \int_0^t z \sin z^2 dz
$$
and 
$$
C'(t) = \int_0^t z \cos z^2 dz.
$$

For our purposes, we care primarily about the properties of the solution as $t'$ goes from $-\infty$ to $+\infty$ (i.e.
the transformation of the orbit as it passes well beyond the resonance).
Thus, we can make use of the identity
$$
\int_0^{\infty} x^m e^{i x^n} dx = \frac{1}{n}\Gamma \left(\frac{1+m}{n}\right) e^{i \pi (1+m)/2 n}
$$
to evaluate
$S(\infty) = C(\infty) = \sqrt{\pi/8}$, $S(-t)=-S(t)$, $C(-t) = -C(t)$ (as in MM11), as
well as $S'(\infty) = 1/2$, $C'(\infty) = 0$, and $S'(-t) = S'(t)$, $C'(-t)=C'(t)$. So,
we find that the evenness of the functions $S'$ and $C'$ mean that the contributions of the
$\dot{\epsilon}$ term to the asymptotic behaviour average out and we recover the same result
as MM11.

Thus, our final expressions are the same as those from MM11,
$$
J_f = J_i + \frac{\pi \epsilon^2}{2 \left| \dot{g}_0 \right|} + \epsilon \sqrt{ \frac{2 \pi J_i}{\left| \dot{g}_0 \right|}} \cos \bar{\omega}
$$
with the equivalent expresson for the inclination
$$
Z_f = Z_i + \frac{\pi \delta^2}{2 \left| \dot{h}_0 \right|} + \delta \sqrt{ \frac{2 \pi Z_i}{\left| \dot{h}_0 \right|}} \cos \Omega.
$$

The expressions for $\epsilon$ and $\delta$ are as above, and the expressions for $\dot{g}_0$ and $\dot{h}_0$ are
\begin{equation}
\frac{dg_0}{dt}=  
 \frac{1}{a} \frac{da}{dt} n  \left[ \frac{1}{8}  \sum_j \frac{m_j}{m_c} \alpha_j^2 \gamma_j
 - \frac{15}{2} \frac{G m_c}{c^2 a} \right],
\end{equation}
and
\begin{equation}
\frac{dh_0}{dt}   = \frac{1}{a} \frac{da}{dt} n \frac{1}{8} \sum_j \frac{m_j}{m_c} \alpha_j^2  \gamma_j
\end{equation}
where $\gamma_j =  b^{(1)}_{3/2}+3 \alpha_j \left[ b^{(0)}_{5/2} - 2 \alpha
 b^{(1)}_{5/2} + b^{(2)}_{5/2} \right]$.

For the 55~Cancri system, the relevant frequencies and couplings are all dominated by 55~Canc~b, with contributions
from 55~Canc~c and relativity contributing at the one percent level. The eigenfunction for the mode that drives the
excitation through the resonance actually contains a stronger contribution (see Figure~\ref{afg}) from 55~Canc~c (by a factor $\sim 5$), but
the greater mass and proximity of 55~Canc~b means that the functions $\epsilon$ and $\delta$ still only contain a
$\sim 5\%$ contribution from 55~Canc~c. Thus, in our discussions, we will refer primarily to excitation by interaction
with 55~Canc~b.

\section{Calibration of the Relativistic Precession and Numerical Tidal Dissipation}
\label{NumCal}

\begin{figure}
\includegraphics[width=84mm]{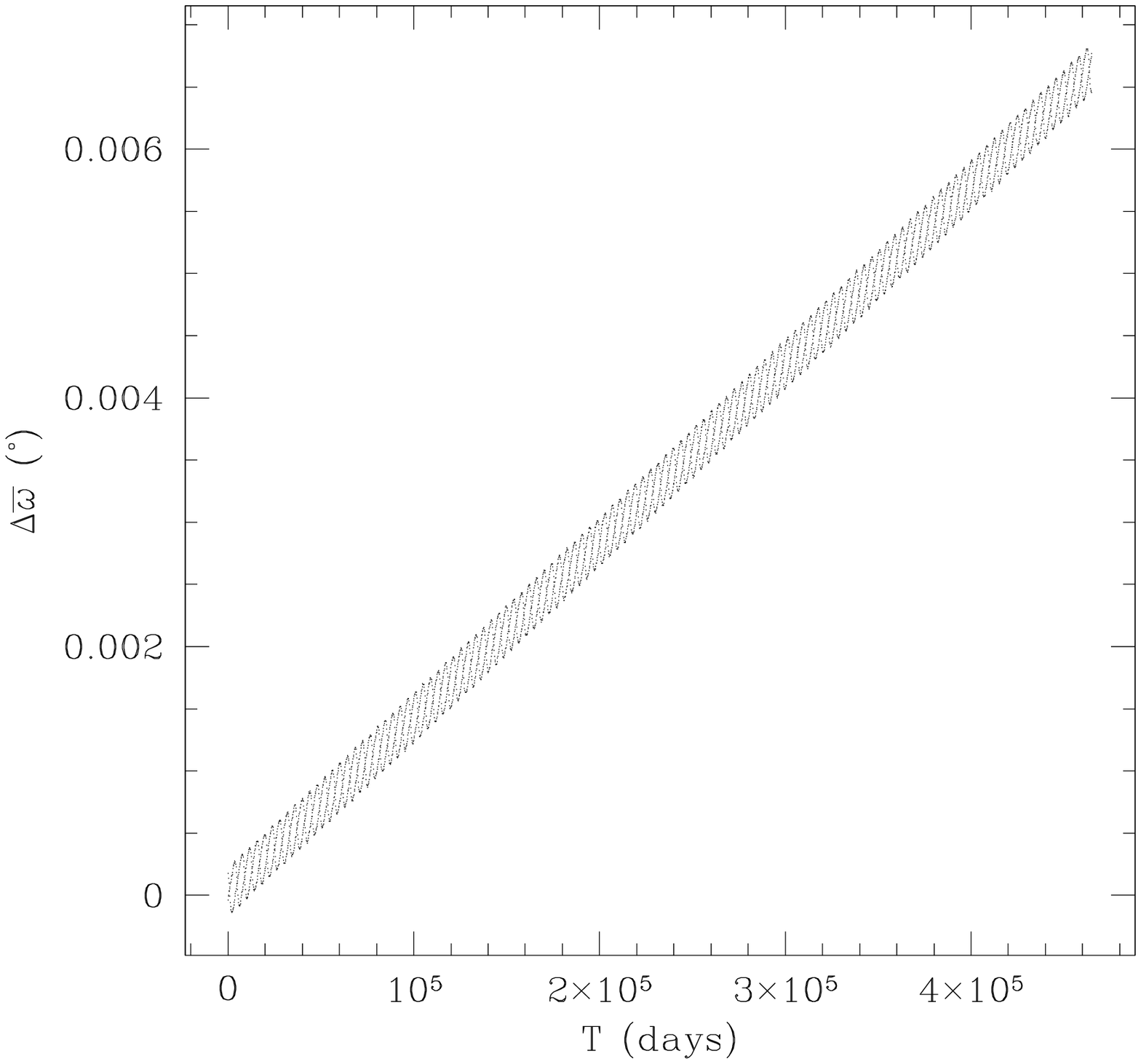}
\caption[RelTest.ps]{The points show the accumulated deviation between the numerically integrated argument of periastron and the
value calculated analytically, for a planet of semi-major axis 0.05~AU, and eccentricity $e=0.2056$. Only a single planet was
included and no tidal damping was included.
\label{RelTest}}
\end{figure}

The Mercury code (Chambers 1999) does not include a prescription for relativistic precession or tidal damping, so we incorporate an
extra user-defined force as follows. The relativistic precession term is standard but can sometimes require special handling for
proper implementation (e.g. Saha \& Tremaine 1994). However, our interests are on sufficiently small scales that our timesteps are
chosen small enough to properly resolve the precession anyway, and a direct implementation with the MVS symplectic algorithm yields
answers of sufficient accuracy. Figure~\ref{RelTest} shows the accumulation of the difference between the numerically calculated
relativistic precession using Mercury and the expected analytic value (calculated for a single planet at 0.05~AU and with no
tidal effects included). The drift of the location of periastron is $\sim 5 \times 10^{-6} \,^{\circ}.yr^{-1}$, well below any
of the precession frequencies of interest in this system (see Figure~\ref{afg}).

 We have previously (Hansen 2010) calibrated tidal interactions based on the constant time-lag
model of Hut (1981) and Eggleton, Kiseleva \& Hut (1998), and so we adopt the force law from Hut (1981). The component
that damps eccentricity is that which acts to dissipate the radial component of the planetary velocity, namely
$$
F = -\alpha \frac{\bf{r}.\bf{\dot{r}}}{r^9} 
$$
where $\bf{r}$ and $\bf{\dot{r}}$ are the heliocentric radius and radial velocity vectors. In Hut's formula, the
constant $\alpha = 9 G M_*^2 k \tau R_p^5$, where $k$ is the apsidal constant, $\tau$ a timescale that encodes the
strength of dissipation, and $R_p$ is the radius of the planet (in which the tides are being dissipated). For the
purposes of calibrating the simulations $\alpha$ is a single number, whose physical relevance will depend on the
relationship between the various physical quantities of which it is composed. In the low eccentricity limit, the
resulting evolution of the eccentricity is an exponential damping, whose characteristic timescale is determined
by $\alpha$. 

\begin{figure}
\includegraphics[width=84mm]{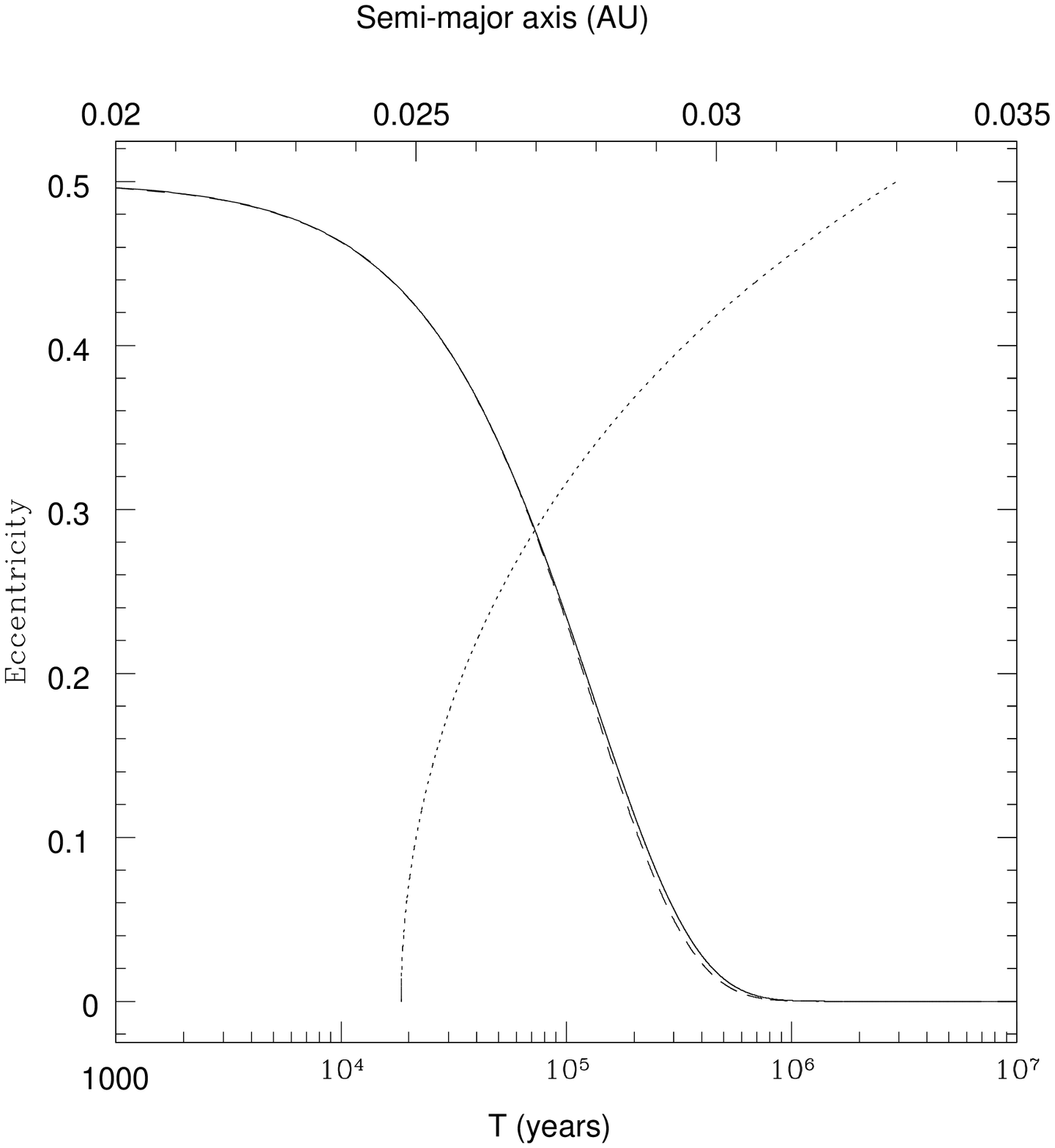}
\caption[Cal.ps]{The solid line shows the evolution of an isolated 55~Canc~e analogue, under the influence of
our tidal dissipation prescription. The plot actually contains a dashed line representing $e = 0.5 {\rm exp}(-t/1.3 \times 
10^5 {\rm yrs})$, but it is hard to see because the fit is excellent. The dotted line relates the eccentricity to
the semi-major axis, given on the upper axis, and demonstrates the conservation of angular momentum during tidal
decay.
\label{Cal}}
\end{figure}

Figure~\ref{Cal} shows the evolution for a 55~Canc~e analogue, begun at 0.033~AU, with eccentricity=0.5 with the
same calibration for $\alpha$ as the simulations described in the main text. This is very well fit by an exponential
decay on a timescale of $1.3 \times 10^5$~years. In the formalism of Hansen (2010), and assuming a planet of
mass $8 M_{\oplus}$ and radius $2 R_{\oplus}$, this corresponds to
$\sigma_p \sim 2.5 \times 10^{-50} g^{-1} cm^{-2} s^{-1}$. This is a considerably stronger level of bulk dissipation
than found for giant planets in that paper, but such a difference is to be expected for a terrestrial-class planet,
and indeed, it is only a factor of 10 larger than the nominal value used by Bolmont et al (2013), based on an
extrapolation from Earth dissipation levels. If cast in terms of the traditional tidal $Q$, it corresponds to $Q_p=13$,
using the expression from Jackson, Greenberg \& Barnes (2008), which is also comparable with an earth-like
dissipation. 

If we adopt a Neptune mass and radius of $20 M_{\oplus}$ and $4 R_{\oplus}$, this would correspond to $Q_p \sim 150$,
which is a little low compared to that observed for Uranus \& Neptune (Banfield \& Murray 1992; Zhang \& Hamilton 2008).

Figure~\ref{Cal} also shows the trajectory of the planet evolution in semi-major axis as a function of eccentricity. The curve
traces out that expected due to the conservation of angular momentum during isolated tidal evolution. It is also worth
noting that the eccentricity properly asymptotes to zero at late times. This can be compared with the late-time behaviour
shown in the full runs in Figure~\ref{Example} and \ref{Example2}, which show some fluctations. The fact that they do
not occur here demonstrates that the effect is a real one, due to residual secular perturbations from the other planets,
and not due to numerical inaccuracies.

\section{Resonant Interference with the Secular Interactions}
\label{Almost}

The numerical experiments demonstrate that the planets 55 Canc~b and c are close enough to the 3:1 resonance that their
resonant interactions may sometimes distort the secular structure of their interactions with 55~Canc~e. Figures~\ref{Merc21_test}
and \ref{Merc22_test} showed how this can manifest itself in the Lagrangian forcing of test particles. A more direct demonstration
can be found in the Fourier decomposition of an integration of the full planetary system. We have performed a fourier transform
of the time series of the eccentricity of 55~Canc~b for each of the two cases shown in those figures,
\begin{equation}
C_k = \sum_{j=0}^{N-1} e_j e^{2 \pi i j k/N}
\end{equation}
and calculated the power spectral density
\begin{equation}
P(f) = \frac{1}{N^2} \left[ \left| C_k \right|^2 + \left| C_{N-k} \right|^2 \right],
\end{equation}
for a time series with a 10 day timestep and $N = 2^{21}$.

\begin{figure}
\includegraphics[width=84mm]{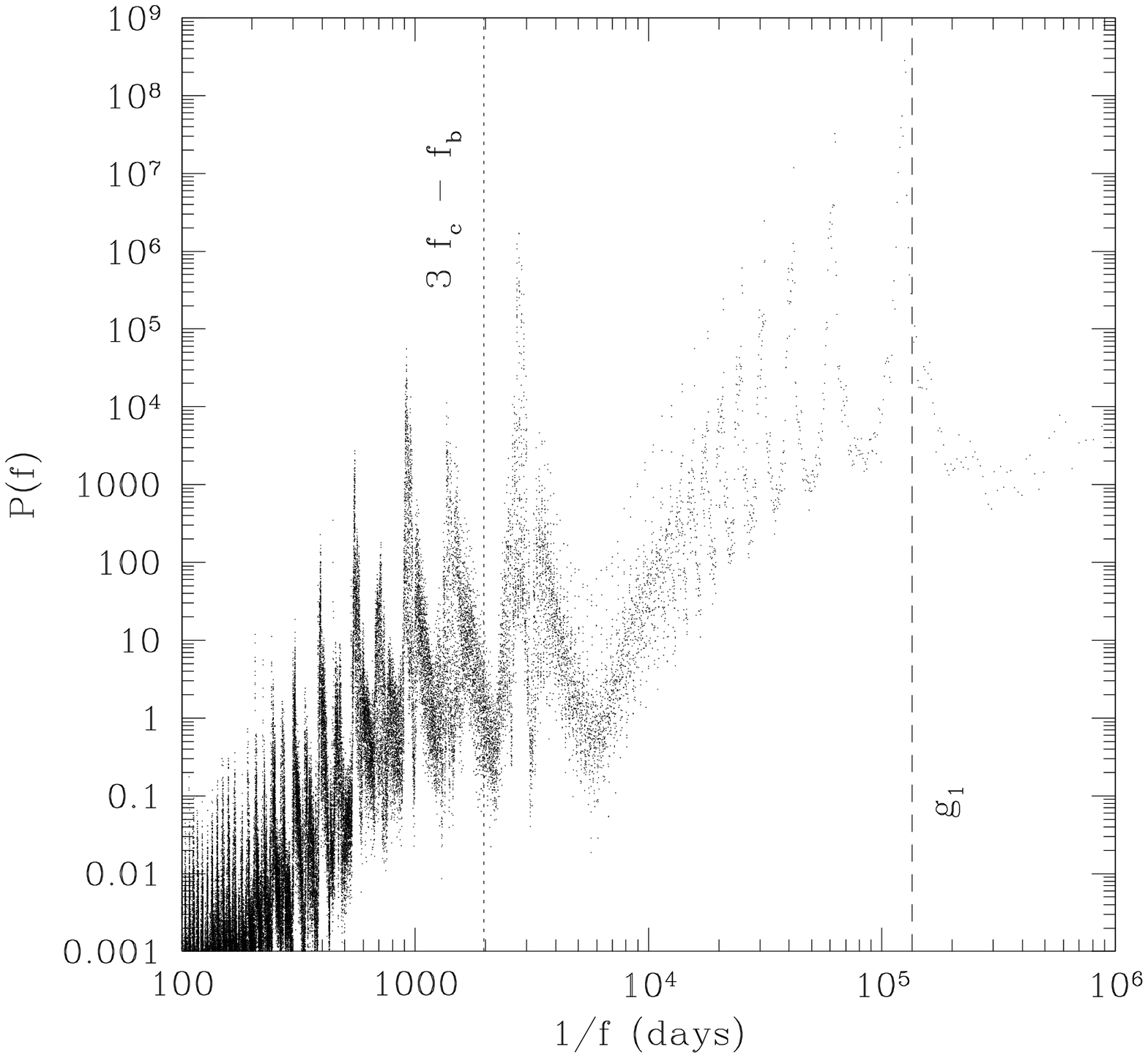}
\caption[C2.ps]{$P(f)$ is the power spectral density of the eccentricity of 55~Canc~b, plotted against the corresponding
period in the time series. The vertical dotted line shows the period associated with the expected circulation frequency
$3 n_c - n_b$ of the planet pair 55~Canc~b and 55~Canc~c. The vertical dashed line is the expected secular frequency $g_1$
based on classical secular perturbation theory.
\label{C2}}
\end{figure}

Figure~\ref{C2} shows the  power spectrum $P(f)$ for the case shown in Figure~\ref{Merc21_test}, which 
conforms to the standard secular evolution we expect. The power spectrum shows the strongest power at the expected secular
frequency $g_1$ (the time series was not long enough to capture the longer secular period) with additional power in the 
neighbourhood of the circulation frequency of the near resonance. There is also a minimum in the power between these two
values, suggesting a clear separation between the resonant and secular phenomena.

\begin{figure}
\includegraphics[width=84mm]{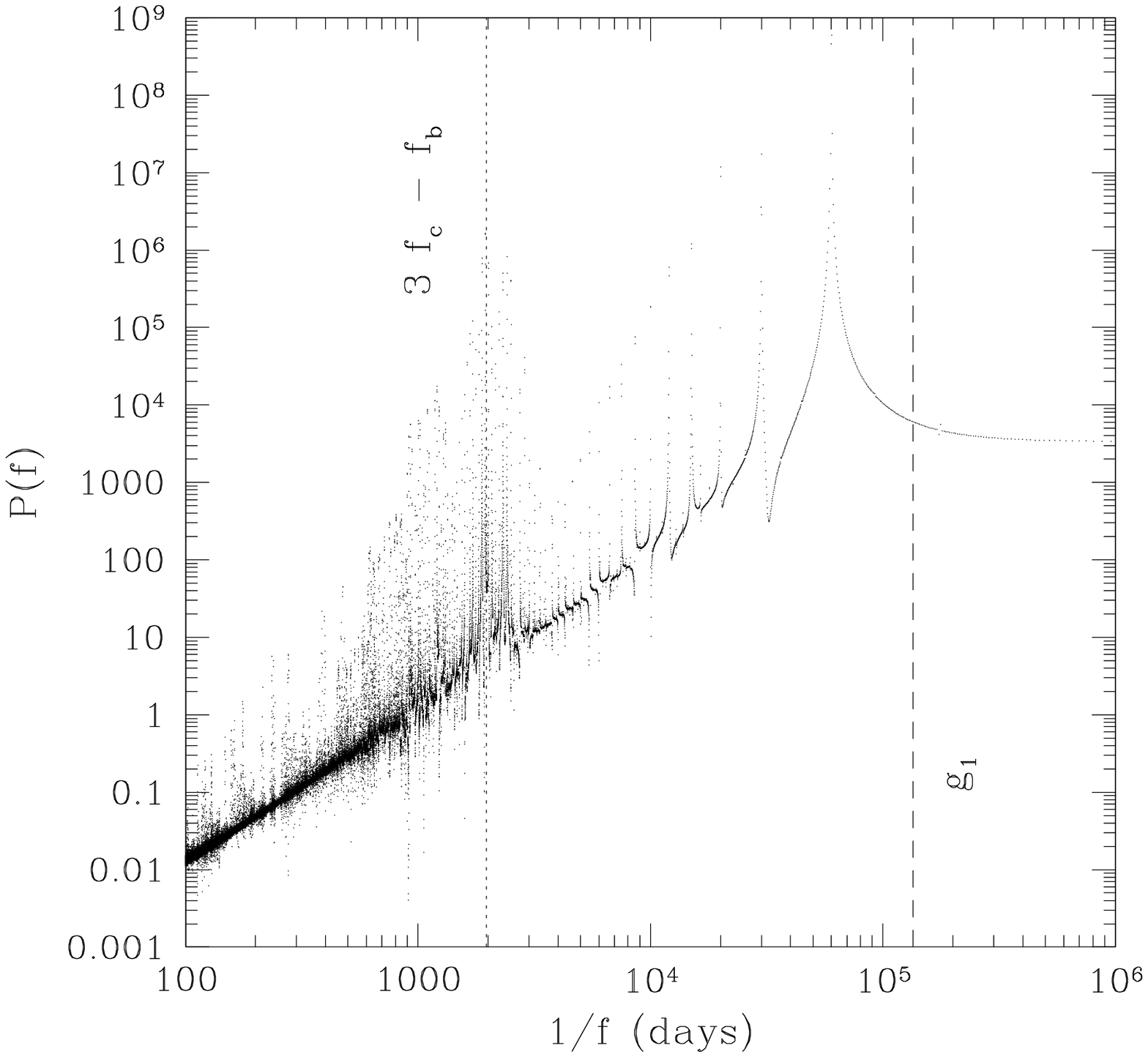}
\caption[C1.ps]{$P(f)$ is once again the power spectral density of the eccentricity of 55~Canc~b, but now for a case
in which the behaviour deviates significantly from the expectations of classical secular theory. In particular, we note
that the period of the dominant secular mode is now substantially shifted from the vertical dashed line. A comparison
with the structure observed in Figure~\ref{C1} demonstrates that the power in the frequency range between the secular
frequency and the circulation frequency is higher, indicating a stronger degree of interaction.
\label{C1}}
\end{figure}

For a different set of initial eccentricities and longitudes of periastron, corresponding to Figure~\ref{Merc22_test}, we find the power spectrum shown in Figure~\ref{C1}.
In this case the secular frequency is shifted to higher values relative to the classical value $g_1$, and more power is found
between this frequency and the higher circulation frequency, indicating a stronger interaction. The shift in frequency of the
secular mode is consistent with the change in the location of the secular resonance observed in Figure~\ref{Merc21_test}.

\begin{figure}
\includegraphics[width=84mm]{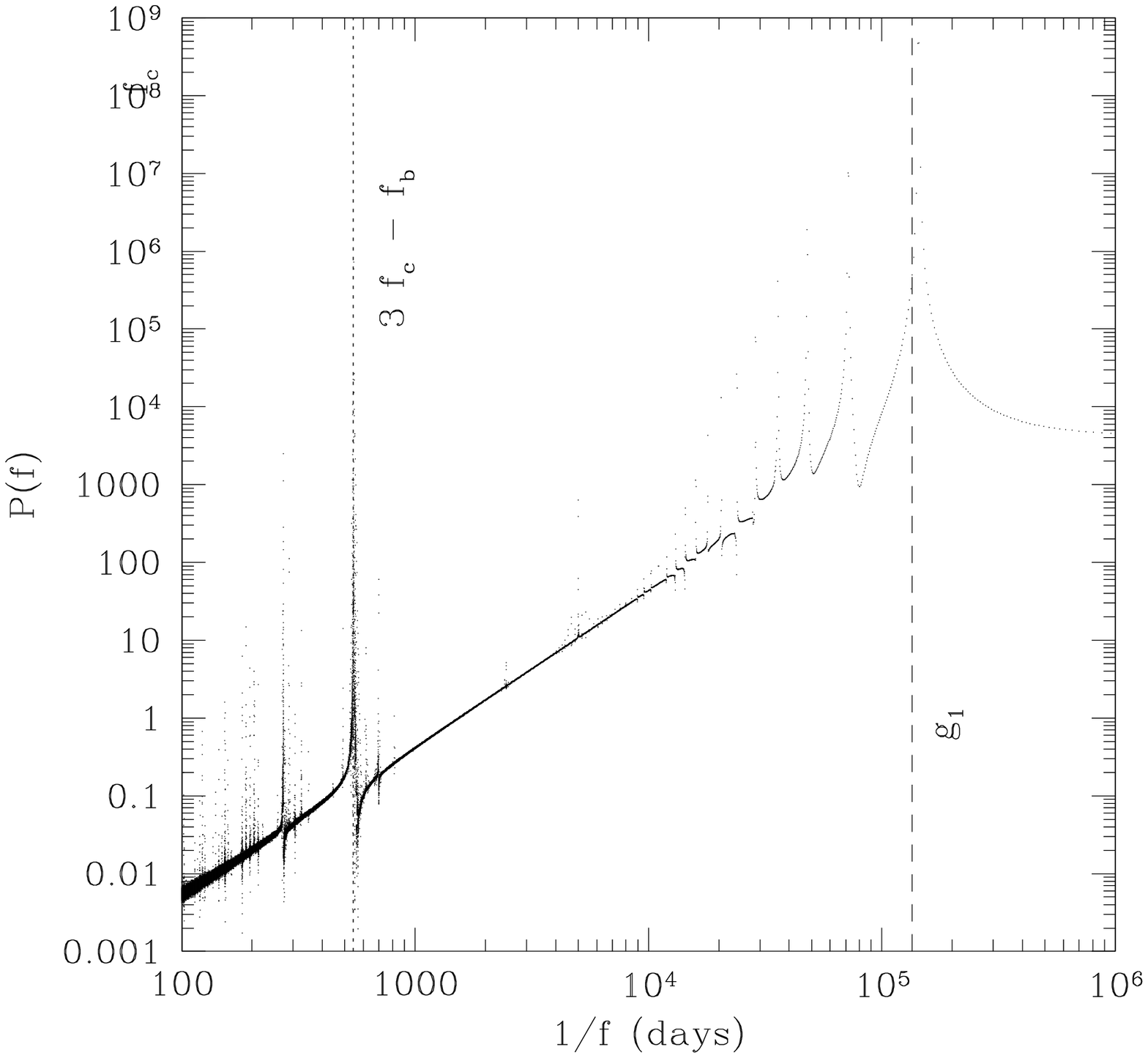}\caption[C0.ps]{
$P(f)$ is once again the power spectral density of the eccentricity of 55~Canc~b, but now for a case in which the semi-major axis of 55~Canc~c has been moved out by 1\%, to reduce the effects of resonant interactions. All other parameters, including initial eccentricities and phases, are the same as in Figure~\ref{C1}. We see here that the distinction between the secular interactions and the circulation of the resonant angle is much clearer, and that the values correspond to the expectatios of classical secular theory.
\label{C0}}
\end{figure}

Could this simply be the case of a set of initial conditions that put almost no power into the mode with frequency $g_1$,
rather than a shift driven by proximity to resonance? Figure~\ref{C0} shows the power spectrum of a simulation run with
exactly the same parameters as in the case of Figure~\ref{C1}, except that the semi-major axis of 55~Canc~c was moved from
0.237~AU to 0.24~AU. We see that this change in 1\% is sufficient to restore the mode frequency to it's expected classical value,
indicating that the observed change in frequency is directly linked to the proximity to resonance.

\begin{table*}
\centering
\begin{minipage}{140mm}
\caption{Constants that define the averaged secular and resonant interactions
of the 55~Canc~b and 55~Canc~c planet pair. \label{Ctab}}
\begin{tabular}{@{}lclclc@{}}
\hline
$C_1$ & 0.263 & $C_4$ & 2.996 & $C_7$ & 1.112 \\
$C_2$ & 0.139 & $C_5$ & -6.133 &  \\
$C_3$ & -2.149& $C_6$ & 1.197 & \\
\hline
\end{tabular}
\end{minipage}
\end{table*}

To gain a qualitative understanding of this phenomenon, let us collect the relevant secular and resonant terms, up
to second order in eccentricity, for the pair 55~Canc~b and 55~Canc~c. The disturbing functions, averaged over short
period terms, are
\begin{eqnarray}
< R_b >=  8.73 \times 10^{-4} \left[ C_0 + C_1 \left( e_b^2 + e_c^2 \right) + \right. \nonumber \\
 \left. C_3 e_b e_c \cos \left( \bar{\omega}_c - \bar{\omega}_b \right) + C_4 e_b^2 \cos \left( \Theta - 2 \bar{\omega}_b \right) 
\right. \nonumber \\  \left. + C_5 e_b e_c \cos \left( \Theta - \bar{\omega}_b - \bar{\omega}_c \right)
+ C_6 e_c^2 \cos \left( \Theta - 2 \bar{ \omega}_c \right) \right] \label{Rb} \\
< R_c >=  4.133 \times 10^{-3} \left[ C_0 + C_1 \left( e_b^2 + e_c^2 \right) + \right. \nonumber \\
 \left. C_3 e_b e_c \cos \left( \bar{\omega}_c - \bar{\omega}_b \right) + C_4 e_b^2 \cos \left( \Theta - 2 \bar{\omega}_b \right) \right. \nonumber \\
 \left. + C_5 e_b e_c \cos \left( \Theta - \bar{\omega}_b - \bar{\omega}_c \right)
+ C_7 e_c^2 \cos \left( \Theta - 2 \bar{ \omega}_c \right)  \right], \label{Rc}
\end{eqnarray}
where the $C_i$ are given in Table~\ref{Ctab} and the difference between $C_6$ and $C_7$ is due to the differing contributions to the
3:1 resonance from the internal and external components of the indirect disturbing function. Far from resonance, the terms in
 the angle $\Theta = 3 \lambda_c - \lambda_b$
will average  to zero and the system will exhibit the traditional secular oscillations. However, if the secular oscillations bring
the planetary eccentricities to low enough values, the resulting increase in the precession rate can sweep the system 
through resonance, inducing a brief libration.

\begin{figure}
\includegraphics[width=84mm]{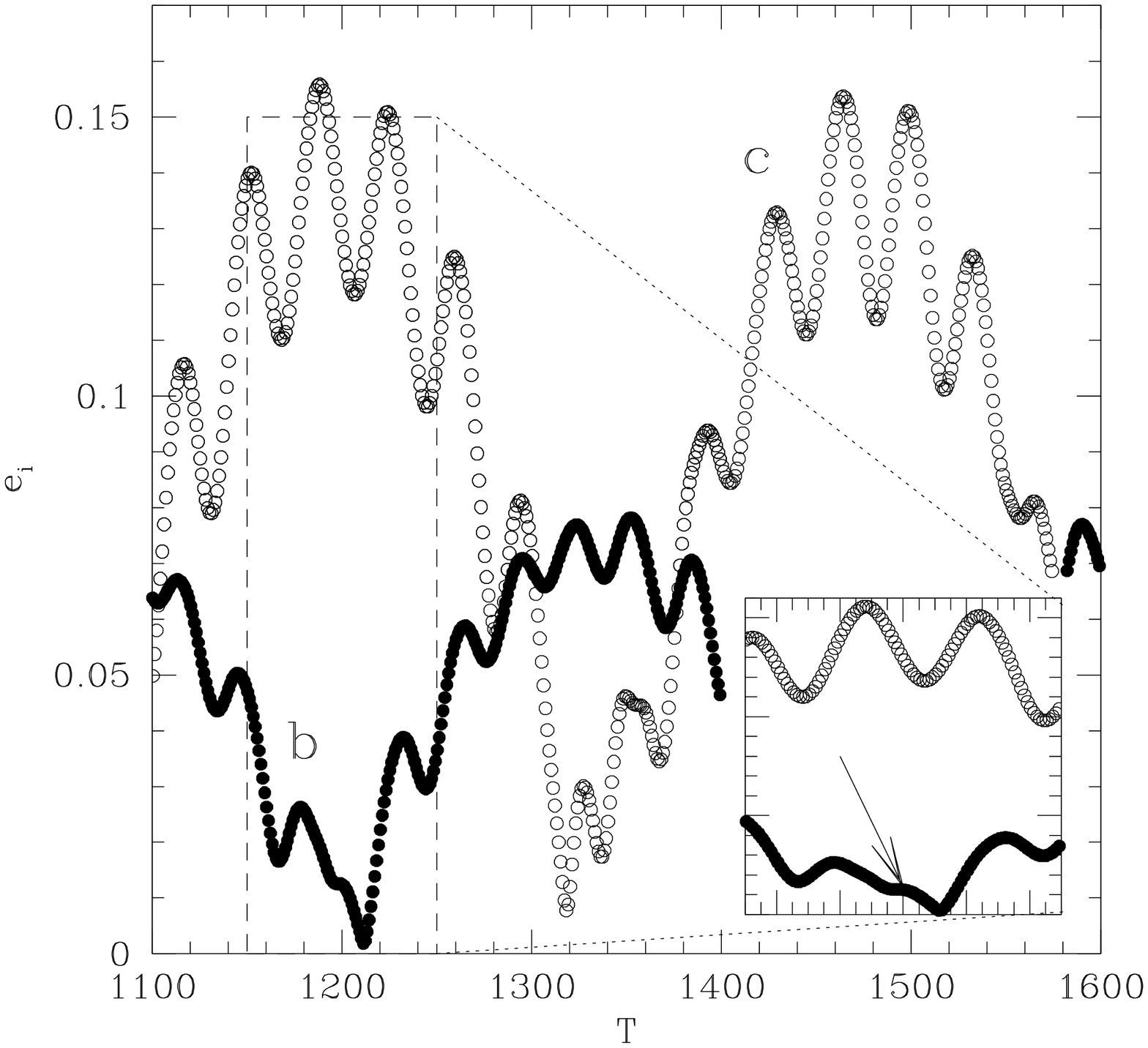}\caption[Et.ps]{The solid points show the eccentricity of 55~Canc~b, integrated
using the Lagrange equations and disturbing functions~(\ref{Rb}) and (\ref{Rc}), while the open circles show 55~Canc~c. The curves
show longer timescale secular oscillations allied to shorter timescale resonant oscillations. The inset shows a
zoom of the passage through a brief libration in the angle $\phi = 3 \lambda_c - \lambda_b - 2 \bar{\omega}_b$, with
the arrow indicating the time of the libration. This corresponds to a change in the phase of the secular oscillation.
\label{Et}}
\end{figure}

We integrate the Lagrange equations based on equations~(\ref{Rb}) and (\ref{Rc}) 
 to describe the evolution of the 55~Canc~b/55~Canc~c system, for a
set of initial conditions that showed the resonant interaction during the full numerical integration. 

\begin{figure}
\includegraphics[width=84mm]{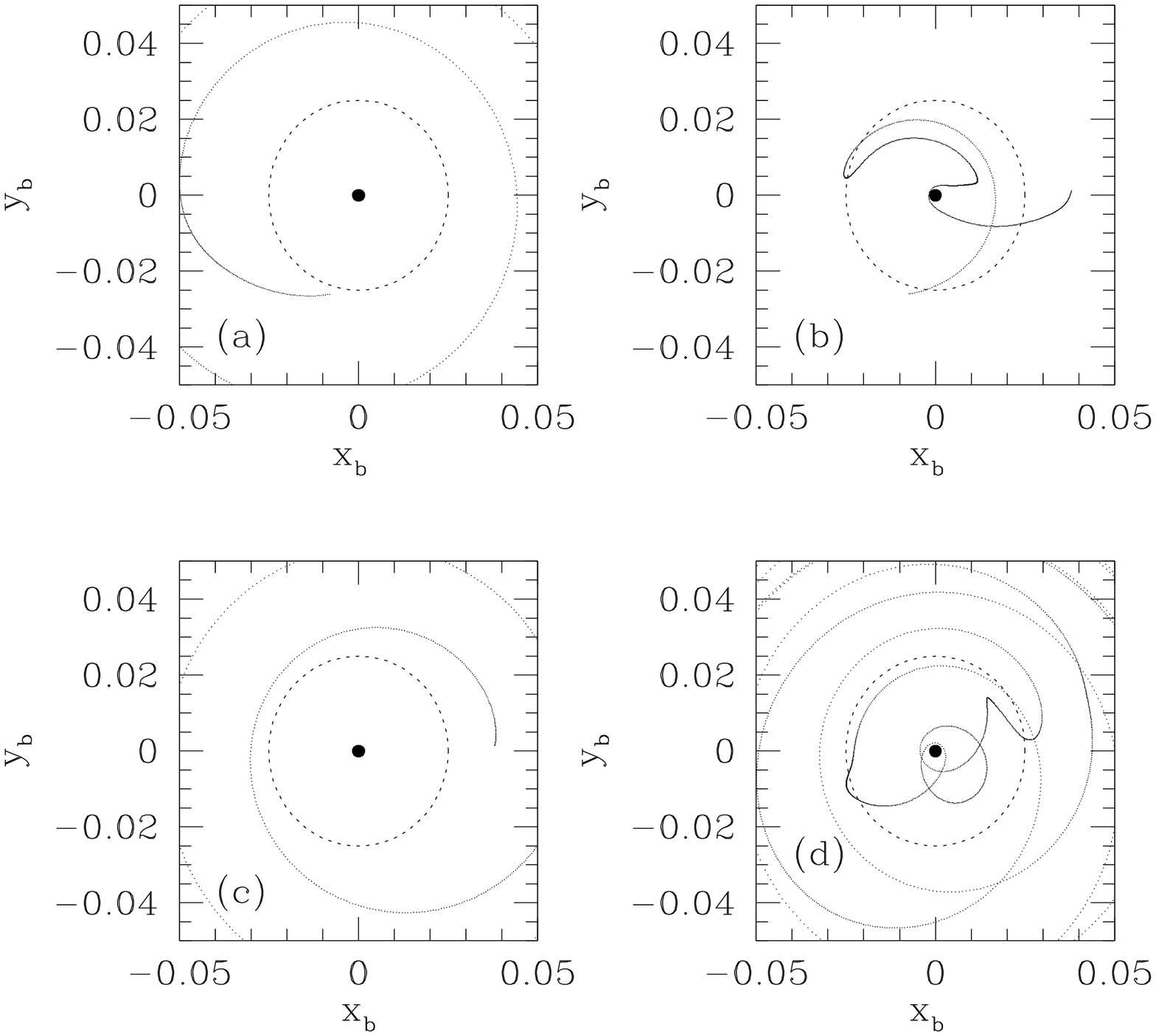}\caption[Full.ps]{Panels a--d indicate a sequence of trajectories for the quantities
$x_b$ and $y_b$ as they pass through an episode of transient libration. The particular interaction show
here is also the one responsible for the evolution during the inset in Figure~\ref{Et}. In particular, panel (a) covers
the time 1100--1160, (b) from 1160--1230, (c) from 1230--1300 and (d) shows the evolution from here through the next
interaction, i.e. from 1300--1600. The dotted circles in each panel have a radius of 0.025 and represent the critical eccentricity
below which one expects an unstable equilbrium point to appear in this resonance, as discussed in the text.
\label{Full}}
\end{figure}

\begin{figure}
\includegraphics[width=84mm]{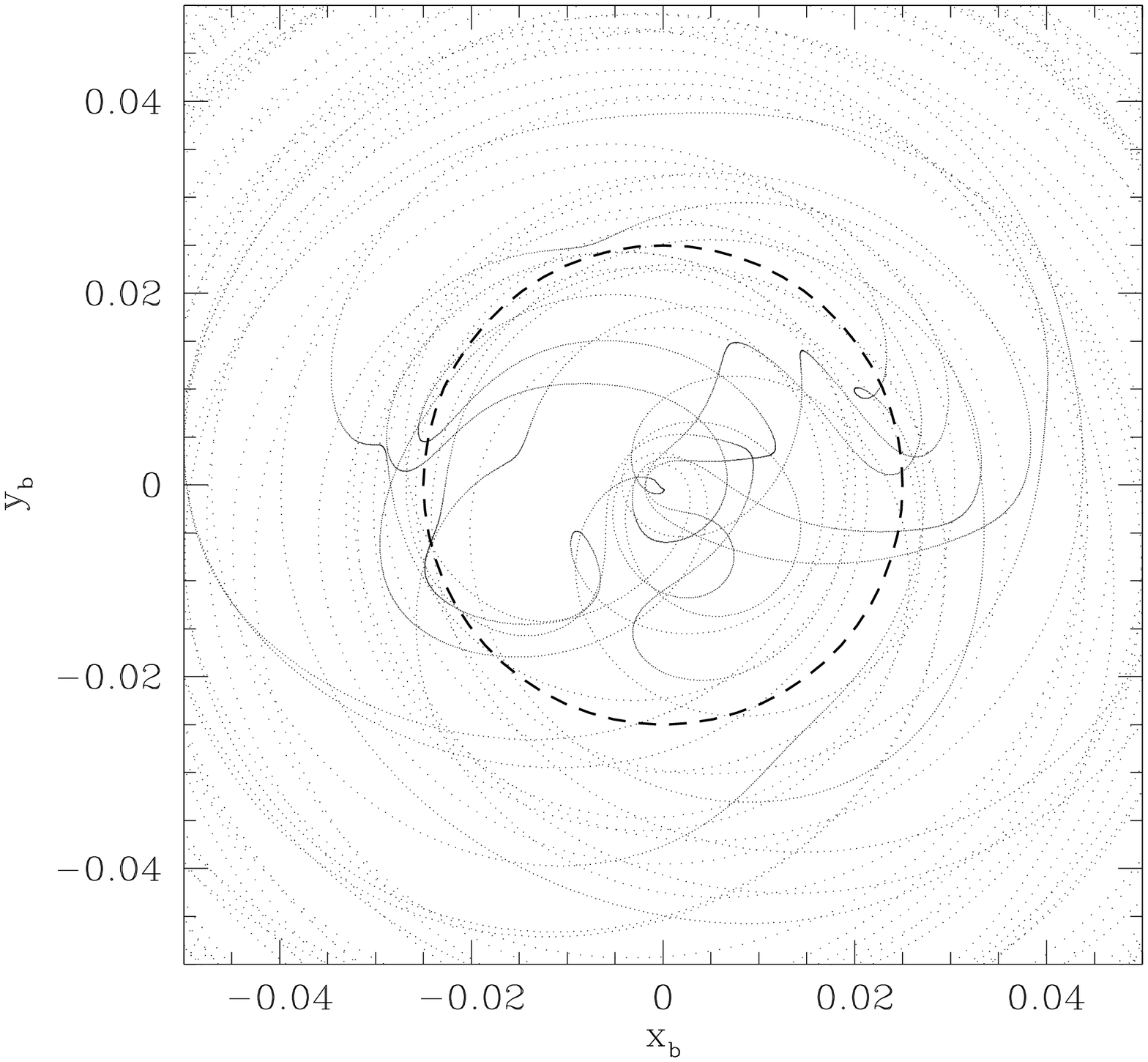}\caption[Trace.ps]{The heavy dashed curve represents the critical eccentricity of 0.025,
and the smaller dots show the trajectory of the system evolution, which is characterised by a simple circulation exterior to
this curve but by a more complicated and chaotic evolution interior to this value, as the system exhibits a resonant interaction.
\label{Trace}}
\end{figure}

As Figure~\ref{Et} demonstrates,
 the evolution that arises  exhibits the characteristic oscillatory behaviour
of secular evolution, with the planetary eccentricities undergoing quasi-sinusoidal oscillations with a range of frequencies.
The most rapid oscillations are associated with the circulation of the resonant argument.
However, we also see the occasional transient librations which act to change the phase of the oscillations and equivalently the 
amplitudes of the oscillations. The inset in Figure~\ref{Et} shows the effect of one of these librations, which acts to essentially apply
a random phase kick to the overall oscillations. Figure~\ref{Full} demonstrates the nature of the deviations from simple
circulation during this interaction. The four panels show a sequence of trajectories in the space 
$x_b = e_b \cos \left(\Theta - 2 \bar{\omega}_b\right)$ and $y_b = e_b \cos\left( \Theta - 2 \bar{\omega}_b\right)$. When the eccentricity gets small, the trajectory exhibits a reversal, which
represents a transition to libration.
 A trajectory for a more lengthy integration is shown in Figure~\ref{Trace}, which demonstrates the presence of a critical eccentricity
below which the circulation transitions to libration.

To understand this, let us extract the resonant Hamiltonian from the relevant part of $<R_b>$, to study the dynamics of
the $e_b^2$ resonance at low $e_b$. Normalising by the mass, 
and defining $\phi_b = \Theta - 2 \bar{\omega}_b$ and $\Delta \bar{\omega} = \bar{\omega}_b - \bar{\omega}_c$, we get the Hamiltonian
\begin{eqnarray}
H=~\left[ C_4 e_b^2 + C_5 e_b e_c \cos \Delta \bar{\omega} + C_6 e_c^2 \cos 2 \Delta \bar{\omega} \right] \cos \phi_b \nonumber \\
+~\left[ C_5 e_b e_c \sin \Delta \bar{\omega} + C_6 e_c^2 \sin 2 \Delta \bar{\omega}\right] \sin \phi_b. \label{Hamb}
\end{eqnarray}

\begin{figure}
\includegraphics[width=84mm]{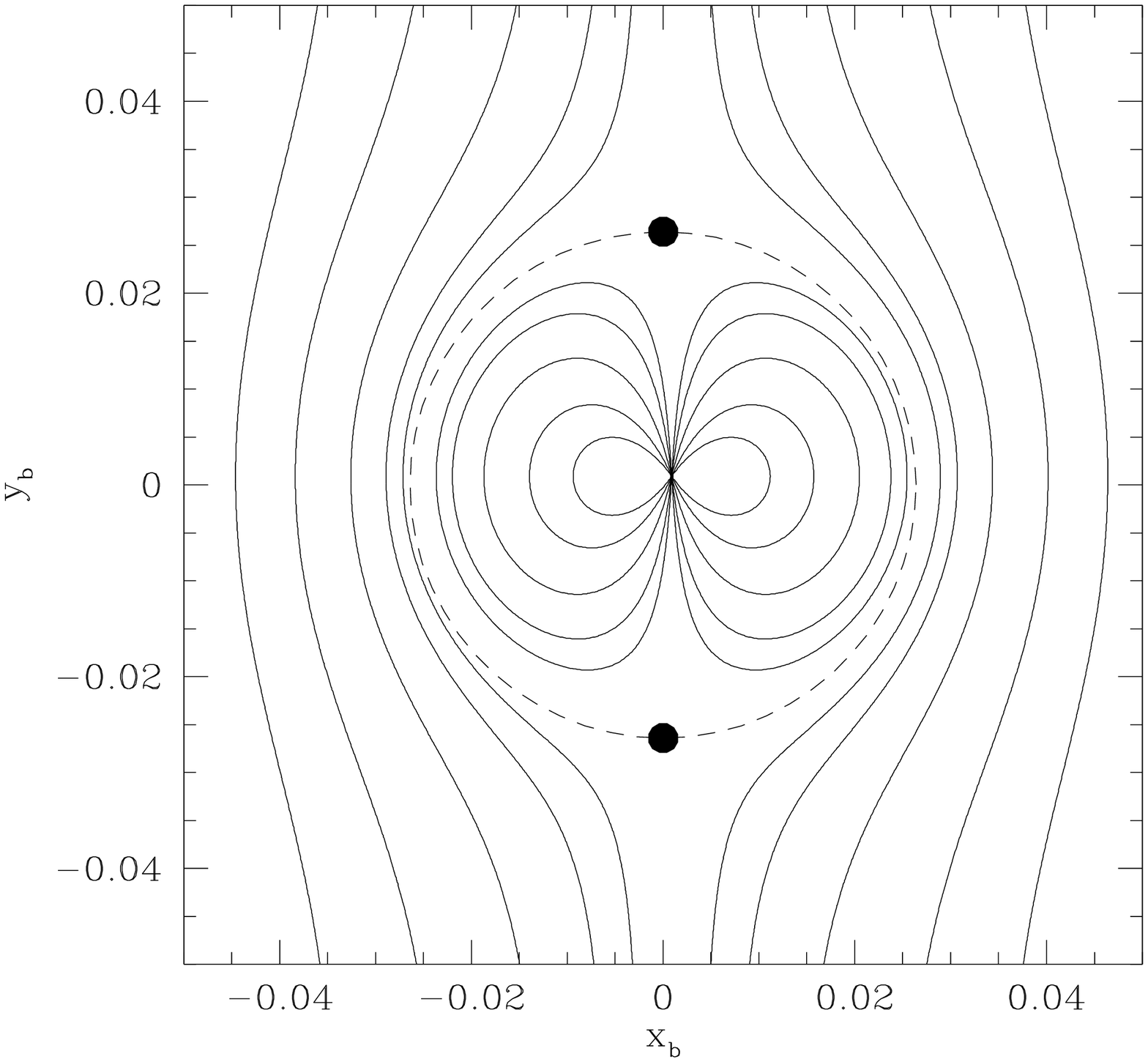}\caption[Ham.ps]{The curves show lines of constant $H_b$, based on equation~(\ref{Hamb}), for the case $e_c=0.12$. The dashed line indicates
a circle of $e_b = 0.026$. The two solid points indicate the equilibrium points on these scales. The sign of $H$ changes upon passage
through the critical $e_b$ at fixed $x_b$, which drives the kinds of reversals of trajectory seen in Figure~\ref{Full}, because the path
has to change from one side of the origin to the other.
\label{Ham1}}
\end{figure}

This Hamiltonian has several equilibrium points, but some are for the case when $e_b$ and $e_c$ are of similar magnitude, which is not
our focus here. In the  
case when $e_b \ll e_c$ and $\bar{\omega}_b=\bar{\omega}_c$, we find two equilibrium points at $x_b=0$ and 
\begin{equation}
\left| y_b \right| = -\frac{e_c}{2 C_4} \left( - C_5 \pm \left[ C_5^2 -4 C_4 C_6 \right]^{1/2}\right) = 0.217\, e_c.
\end{equation}

This value of $y_b$  defines a characteristic value of $e_b$ at which the circulation of the resonant argument switches
over to libration.
Figure~\ref{Ham1} shows the curves of constant $H_b$ for the case $e_c=0.12$ (corresponding to the value during the resonance passage
shown in the inset of Figure~\ref{Et}), which yields the critical value of $e_b = 0.026$ (shown as
a dashed circle).
 When the secular oscillations bring $e_b$ below this value, the increase in secular precession sweeps the system into resonance, and 
the trajectory only returns to simple circulation when the oscillations bring $e_b$ to larger values again. The behaviour is similar when
the periapses are not aligned, but the locations of the equilibrium points rotate in $x_b$ and $y_b$. The dramatic behaviour seen in Figures~\ref{Full} and \ref{Trace} are thus manifestations of the passage through the resonance as the rate of secular precession increases while
$e_b$ decreases. 

Similar considerations describe the behaviour of the $e_c^2$ resonance, when $e_c \ll e_b$. Thus, although the nominal timescales for
secular oscillations and mean motion librations are separated by almost two orders of magnitude, the excursions to low eccentricities
in the secular oscillations are sufficient to enhance the precession rates to sweep the system through the resonance on an intermittent
basis. The resulting interference with the secular oscillations appears to shift the frequencies sufficiently to make a measureable 
difference in the resonant forcing of some of our model systems. However, it appears as though this subset has less likelihood of generating
observable 55~Canc~e analogues, as most of the inner planets in these systems end up hitting the star (\S~\ref{NumTides}).

\end{document}